\documentclass[useAMS,usenatbib,onecolumn]{mn2e}

\usepackage{graphicx}

\font\syvec=cmbsy10                        
\font\gkvec=cmmib10                         
\def\bnabla{\hbox{{\syvec\char114}}}       
\def\bgamma{\hbox{{\gkvec\char13}}}        
\def\bxi{\hbox{{\gkvec\char24}}}           

\def\baromr{\bar{\omega}_r}
\def\baromt{\bar{\omega}_\theta}
\def\bark{\bar{\kappa}_0}
\def\bars{\bar{\sigma}_0}

\def\spose#1{\hbox to 0pt{#1\hss}}
\def\lta{\mathrel{\spose{\lower 3pt\hbox{$\mathchar"218$}}
     \raise 2.0pt\hbox{$\mathchar"13C$}}}
\def\gta{\mathrel{\spose{\lower 3pt\hbox{$\mathchar"218$}}
     \raise 2.0pt\hbox{$\mathchar"13E$}}}

\title[Oscillation modes of relativistic slender tori]{Oscillation modes of
relativistic slender tori}
\author[O. M. Blaes, P. Arras, and P. C. Fragile]{O. M. Blaes$^{1}$\thanks{E-mail:
blaes@physics.ucsb.edu (OMB); arras@kitp.ucsb.edu (PA); fragilep@cofc.edu
(PCF)}, P. Arras$^{2}$\footnotemark[1] and P. C.
Fragile$^{3}$\footnotemark[1]\\
$^{1}$Department of Physics, University of California, Santa Barbara,
CA 93106, USA\\
$^{2}$Kavli Institute for Theoretical Physics, Kohn Hall,
University of California, Santa Barbara, CA 93106, USA\\
$^{3}$Department of Physics and Astronomy, College of Charleston, 58
Coming Street, Charleston, SC 29424, USA}
\begin{document}

\date{Accepted ---. Received ---; in original form ---}

\pagerange{\pageref{firstpage}--\pageref{lastpage}} \pubyear{2005}

\maketitle

\label{firstpage}

\begin{abstract}
Accretion flows with pressure gradients permit the existence of standing waves
which may be responsible for observed quasi-periodic oscillations (QPO's) in
X-ray binaries.  We present a comprehensive treatment of the linear modes of a
hydrodynamic, non-self-gravitating, polytropic slender torus, with arbitrary
specific angular momentum distribution, orbiting in an arbitrary axisymmetric
spacetime with reflection symmetry.  We discuss the physical nature of the
modes, present general analytic expressions and illustrations for those which
are low order, and show that they can be excited in numerical simulations
of relativistic tori.  The mode oscillation spectrum simplifies dramatically
for near Keplerian angular momentum distributions, which appear to be generic
in global simulations of the magnetorotational instability.  We discuss our
results in light of observations of high frequency QPO's, and point out the
existence of a new pair of modes which can be in an approximate 3:2 ratio for
arbitrary black hole spins and angular momentum distributions, provided the
torus is radiation pressure dominated.  This mode pair consists of the
axisymmetric vertical epicyclic mode and the lowest order axisymmetric
breathing mode.
\end{abstract}

\begin{keywords}
accretion, accretion discs -- black hole physics -- relativity --
X-rays: binaries.
\end{keywords}

\section{Introduction}

Observations of quasiperiodic oscillations (QPO's) in the X-ray light curves of
black hole and neutron star X-ray binaries have motivated the theory
of relativistic ``diskoseismology'', the study of hydrodynamic
oscillation modes of geometrically thin accretion discs (see
\citealt{wag99} and \citealt{kat01} for reviews).  Standing waves
with discrete frequencies can exist in such discs because the radial
profiles of general relativistic test particle oscillation frequencies
can create finite regions where modes are trapped.
Because the disc is considered to be geometrically thin,
radial pressure gradients are usually negligible in the
equilibrium structure of the accretion flow, which is presumed to consist
of fluid rotating on nearly circular geodesics.

However, radial pressure gradients themselves can permit the existence of
discrete modes because they produce a disc, or torus, of finite extent.
At least in the case of black hole X-ray binaries, QPO's are only observed
in the ``hard'' and ``steep power law'' spectral/variability states
(e.g. \citealt{mcc05}), states in which the flow is not solely composed
of a standard geometrically thin accretion disc.  It is therefore conceivable
that radial pressure gradients may be important in trapping modes, and
various groups have recently explored the possibility that accretion tori
may explain observed QPO's, e.g. \citet{gia04} for low frequency QPO's,
\citet{rez03a,klu04,lee04} for high frequency QPO's.  Geometrically thick
tori may form in stellar collapse, and modes of dense tori around black holes
have also been suggested as a detectable source of gravitational waves (e.g.
\citealt{zan03}).

A complete analysis of the spectrum of modes in tori has not yet been done
except in the limiting case of constant specific angular momentum, slender
tori in a Newtonian point mass gravitational potential \citep{bla85}.  In this
paper we greatly generalize that early work and present a rather complete
analysis of the oscillation modes of non-self-gravitating, polytropic,
hydrodynamic tori with small cross-section and arbitrary specific angular
momentum distributions.  The torus equilibria are assumed
to be stationary and axisymmetric, and orbit in an arbitrary background
spacetime which is itself axisymmetric and possesses reflection symmetry.
Although limited to slender tori, our work here could be numerically extended
to thicker tori by tracking all the mode frequencies.  Armed with a complete
understanding of all the mode frequencies and their physical properties, it
should be possible to better identify what modes might be responsible for
observed QPO's.

An analysis of torus oscillation modes is also useful as a code check for
global numerical simulations of accretion flows.  In part because of their
finite spatial resolution, these simulations generally use tori rather
than geometrically thin discs as initial conditions (e.g.
\citealt{dev03,gam03}).  It is noteworthy that global simulations of
non-radiative, magnetized accretion flows often produce small, albeit highly
variable, hot inner tori near the innermost stable circular orbit
\citep{haw02}.

This paper is organized as follows.  In section 2 we derive the basic equations
describing the equilibria and linear perturbations of relativistic slender
tori.  We then make some general remarks about the types of modes and
their classification in section 3.  In section 4 we present complete
solutions for all the modes of relativistic fluid tori in four special
cases where this can be done analytically.  Then in section 5 we show how
the low order modes of the general torus can be derived exactly, and also
present numerical solutions for higher order modes.  In section 6 we compare
these analytic results with numerical simulations of oscillating hydrodynamic
tori, and we finally summarize our conclusions in section 7.  Readers who
are primarily interested in applications to QPO's should focus on sections
5 and 7.  There we identify a robust new candidate for the modes comprising
the observed 3:2 ratio of high frequency QPO's in black hole X-ray binaries.
We apply our analytic results for these mode frequencies to GRO~J1655-40 in
Figure 9.

We remind the reader that slender hydrodynamic
tori are violently unstable to the Papaloizou-Pringle instability
\citep{pap84}.  Perhaps even more important, in the presence of a weak
magnetic field, they are unstable to the development of magnetohydrodynamical
turbulence due to the magnetorotational instability \citep{bal98}.  Whether
or not the hydrodynamic modes we discuss here can survive these instabilities
or even exist is a problem which still needs to be investigated, and we
discuss this briefly in section 7.

\section{Setting Up the Problem}

We consider an ideal, hydrodynamic flow in a stationary, axisymmetric
spacetime with reflection symmetry.  The line element may be expressed as
\begin{equation}
ds^2=g_{tt}dt^2+2g_{t\phi}dtd\phi+g_{rr}dr^2+g_{\theta\theta}d\theta^2
+g_{\phi\phi}d\phi^2.
\end{equation}
We take the metric to have signaure $(-+++)$.  Because of the assumed
symmetries, the metric coefficients $g_{\mu\nu}$ depend on $r$ and $\theta$
only, and remain unchanged under the transformation
$\theta\rightarrow\pi/2-\theta$.

We will restrict consideration throughout this paper to flows which are
isentropic, and in fact polytropic.  Hence the internal energy per unit
volume $e$ (including rest mass), pressure $p$, and rest mass density $\rho$,
all measured in the fluid rest frame, are related by
\begin{equation}
e=np+\rho\,\,\,\,\, {\rm and}\,\,\,\,\, p=K\rho^{1+1/n}.
\label{eqpolytrope}
\end{equation}
Here $n$ is the polytropic index and $K$ is a constant.  We use units where
$G=c=1$ here and throughout this paper.

\subsection{Equilibrium}

We consider an equilibrium, axisymmetric torus in a state of pure rotation.
The only nonzero components of the fluid four-velocity are then $u^t$ and
$u^\phi$.  The angular velocity measured by an observer
at infinity is $\Omega=u^\phi/u^t$, and the normalization of the four-velocity
$u^\alpha u_\alpha=-1$ requires
\begin{equation}
A\equiv u^t=\left(-g_{tt}-2\Omega g_{t\phi}-\Omega^2 g_{\phi\phi}\right)^{-1/2}.
\end{equation}
Other useful quantities are the specific energy ${\cal E}$ and specific
angular momentum $\ell$ of a given fluid element:
\begin{equation}
{\cal E}\equiv-u_t = \left(-g^{tt}+2\ell g^{t\phi}-\ell^2 g^{\phi\phi}
\right)^{-1/2}\,\,\,\,\,{\rm and}\,\,\,\,\, \ell\equiv-{u_\phi\over u_t}.
\end{equation}

The spatial $r-\theta$ structure of the torus is given entirely by the
relativistic Euler equation, which in this case may be written
\begin{equation}
{1\over2}{\cal E}^2\left(\bnabla g^{tt}-2\ell\bnabla g^{t\phi}+\ell^2
\bnabla g^{\phi\phi}\right)=-{1\over e+p}\bnabla p.
\label{eqeulerequil}
\end{equation}
Here the operator $\bnabla$ may be taken to be either $\partial/\partial r$
or $\partial/\partial\theta$.

In the equatorial plane inside the torus is a special radius $r_0$ where the
pressure has a maximum and fluid orbits on a circular geodesic.  From equation
(\ref{eqeulerequil}), the ``Keplerian'' specific angular momentum $\ell_0$ at
this radius satisfies
\begin{equation}
\left(g^{tt}_{,r}-2\ell_0g^{t\phi}_{,r}+\ell_0^2g^{\phi\phi}_{,r}\right)_0=0,
\label{eqell0}
\end{equation}
where subscript zero refers to the pressure maximum.

It is convenient to introduce an effective potential
\begin{equation}
{\cal U}=g^{tt}-2\ell_0g^{t\phi}+\ell_0^2g^{\phi\phi}.
\end{equation}
A test particle with specific angular momentum $\ell_0$ orbits on a circular
geodesic at an extremum of ${\cal U}$.  This extremum is of course located
at the pressure maximum $r_0$.  Linear perturbations of a test particle about
this circular geodesic lead to radial and vertical epicyclic oscillations.
The squares of the corresponding angular frequencies, as measured by an
observer at infinity, are given by
\begin{equation}
\omega_r^2={{\cal E}_0^2\over A_0^2}\left({1\over2g_{rr}}{\partial^2{\cal U}
\over\partial r^2}\right)_0\,\,\,\,\,{\rm and}\,\,\,\,\,
\omega_\theta^2={{\cal E}_0^2\over A_0^2}\left({1\over2g_{\theta\theta}}
{\partial^2{\cal U}\over\partial \theta^2}\right)_0,
\end{equation}
respectively \citep{abr05}.

Given our polytropic
equation of state (\ref{eqpolytrope}), the pressure and density can be
expressed in terms of an auxiliary function $f(r,\theta)$:
\begin{equation}
p=p_0f^{n+1}\,\,\,\,\,{\rm and}\,\,\,\,\, \rho=\rho_0f^n.
\end{equation}
Consider now a slender torus in which $p_0/\rho_0$ is very small.
By introducing local coordinates measured from the equilbrium point,
\begin{equation}
x\equiv g_{rr0}^{1/2}\left({r-r_0\over r_0}\right)\,\,\,\,\,{\rm and}\,\,\,\,\,
y\equiv g_{\theta\theta0}^{1/2}\left({\pi/2-\theta\over r_0}\right),
\end{equation}
Abramowicz et al. (2005) expanded equation (\ref{eqeulerequil}) to show that
\begin{equation}
f=1-{1\over\beta^2}\left\{\left[\bar{\omega}_r^2-{{\cal E}_0^2\over A_0^2
\Omega_0^2\ell_0}\left({g^{tt}_{,r}-\ell_0g^{t\phi}_{,r}\over g_{rr}}{\partial
\ell\over\partial r}\right)_0\right]x^2+\bar{\omega}_\theta^2y^2\right\}.
\label{eqfabram}
\end{equation}
Here
\begin{equation}
\beta^2\equiv{2(n+1)p_0\over\rho_0A_0^2r_0^2\Omega_0^2}
\label{eqbeta}
\end{equation}
is a dimensionless parameter which determines the thickness of the torus.
The slender torus limit corresponds to $\beta\rightarrow0$.  The quantities
$\bar{\omega}_r$ and $\bar{\omega}_\theta$ are the epicyclic frequencies
scaled by the angular velocity at the pressure maximum $\Omega_0$.

It turns out that the equilibrium function $f$ given by equation
(\ref{eqfabram}) can be written more simply and physically.  Following
\citet{seg75}, define the spatial vector
\begin{equation}
\bgamma=A{\cal E}^3\left[(1+\ell\Omega)\bnabla g^{t\phi}-\Omega\bnabla g^{tt}
-\ell{\bnabla}g^{\phi\phi}\right]
\label{eqgamma}
\end{equation}
For barotropic configurations such as the polytropic tori we are considering
here, surfaces of constant $\Omega$ and $\ell$ coincide \citep{abr71}.
By taking the curl of equation (\ref{eqeulerequil}), it is straightforward to
show that the vector $\bgamma$ is then perpendicular to these level surfaces,
\begin{equation}
\bgamma\times\bnabla\ell=0.
\label{eqgammacrossell}
\end{equation}
Even more importantly, the quantity
\begin{equation}
\kappa^2\equiv{1\over A^2}\bgamma\cdot\bnabla\ell={1\over A^2}
\left({\gamma_r\over g_{rr}}
{\partial\ell\over\partial r}+{\gamma_\theta\over g_{\theta\theta}}
{\partial\ell\over\partial\theta}\right)
\label{eqkappa2}
\end{equation}
represents the characteristic squared frequency (as measured by an observer
at infinity) of inertial oscillations in the fluid due to the presence of
an equilibrium specific angular momentum gradient.  We show this explicitly
later in this paper, but the result is not surprising given that
$\bgamma\cdot\bnabla\ell$ is the quantity that appears in relativistic
generalizations of the Rayleigh and H{\o}iland local stability criteria
\citep{seg75}.  At the pressure maximum, equations (\ref{eqell0}),
(\ref{eqgamma}), and (\ref{eqkappa2}) imply
\begin{equation}
\kappa_0^2\equiv\bar{\kappa}_0^2\Omega_0^2={{\cal E}_0^2\over\ell_0A_0^2}
\left({g^{tt}_{,r}-\ell_0g^{t\phi}_{,r}\over g_{rr}}{\partial\ell\over
\partial r}\right)_0.
\label{eqkappa02}
\end{equation}
Hence equation (\ref{eqfabram}) may be written as
\begin{equation}
f=1-\left(\bar{\omega}_r^2-\bar{\kappa}_0^2\right)\bar{x}^2
-\bar{\omega}_\theta^2\bar{y}^2,
\label{eqf}
\end{equation}
where $\bar{x}\equiv x/\beta$ and $\bar{y}\equiv y/\beta$.

\subsection{Perturbations}

The equations describing the dynamics of linear perturbations of the torus
may be derived by linearizing the relativistic continuity equation,
\begin{equation}
\nabla_\alpha(\rho u^\alpha)=0,
\end{equation}
the relativistic Euler equation,
\begin{equation}
u^\beta\nabla_\beta u^\alpha=-{u^\alpha u^\mu + g^{\alpha\mu}\over e+p}
\nabla_\mu p
\end{equation}
the equation of state (\ref{eqpolytrope}), and the normalization of the
four-velocity, about the equilibrium discussed
above.  Because the equilibrium is stationary and axisymmetric, we choose all
the perturbation variables to vary with $t$ and $\phi$ according to
$\exp[i(m\phi-\omega t)]$.  \citet{ips92} have shown that the
resulting perturbation equations can then be combined into a single, second
order partial differential equation in terms of a single scalar perturbation
variable.  In order to follow the notation of recent work on oscillations of
tori \citep{abr05}, we define this variable slightly differently
as
\begin{equation}
W\equiv-{\delta p\over\rho A\sigma},
\label{eqwdef}
\end{equation}
where $\sigma\equiv\omega-m\Omega$.  We denote this scalar variable by $W$,
as this was the notation first used in the original Newtonian study of
global instabilities of tori \citep{pap84}.

The perturbation equation for tori could be derived by changing variables
and taking the appropriate limit of the general equation of \citet{ips92},
but we sketch the direct derivation here in order to include some
intermediate results which will be useful.  The direct derivation is greatly
aided by noting that equations (\ref{eqeulerequil}) and (\ref{eqgamma})
imply
\begin{equation}
-{A\sigma W\over e+p}\bnabla p+\bnabla(A\sigma W)=A\sigma\bnabla W
+W\sigma\left[-{\ell\over{\cal E}}\bgamma+{\cal E}(A^2\Omega+g^{t\phi})
\bnabla\ell\right]
+Wm\left[{1-\ell\Omega\over{\cal E}}\bgamma+{\cal E}(\Omega g^{t\phi}
-g^{\phi\phi})\bnabla\ell\right].
\end{equation}
Using this and equation (\ref{eqgammacrossell}), it is straightforward to
show that the $r$ and $\theta$ components of the Eulerian four-velocity
perturbation are given by
\begin{equation}
\delta{\bf u}={i\rho\over e+p}\bnabla W+{i\over A^2(\sigma^2-\kappa^2)}
\left({\rho\over e+p}\right)\Biggl\{\bgamma(\bnabla\ell\cdot\bnabla W)+
\sigma A {\cal E} W\left[\sigma(A^2\Omega+g^{t\phi})+m(\Omega g^{t\phi}
-g^{\phi\phi})\right]\bnabla\ell\Biggr\}.
\label{eqdui}
\end{equation}
The azimuthal component is given by
\begin{equation}
\delta u_\phi={\rho\over e+p}A{\cal E}(\sigma\ell+m\Omega\ell-m)W
+{{\cal E}^2\over A(\sigma^2-\kappa^2)}\left({\rho\over e+p}\right)
\bnabla\ell\cdot\Biggl\{A\sigma\bnabla W+
W{\cal E}\left[\sigma(A^2\Omega+g^{t\phi})+m(\Omega g^{t\phi}
-g^{\phi\phi})\right]\bnabla\ell\Biggr\}
\label{eqduphi}
\end{equation}
These equations are completed by the linearized continuity equation,
\begin{equation}
i{\sigma^2A^2\rho^2n\over(n+1)p}W-i(\sigma\ell+m\Omega\ell-m)\rho
(g^{\phi\phi}-g^{t\phi}\Omega)\delta u_\phi
+{1\over(-g)^{1/2}}{\partial\over\partial r}[(-g)^{1/2}\rho g^{rr}
\delta u_r]
+{1\over(-g)^{1/2}}{\partial\over\partial\theta}[(-g)^{1/2}\rho
g^{\theta\theta}\delta u_\theta]=0.
\label{eqdcont}
\end{equation}
So far, these equations are exact.  In the slender torus limit, $r$ and
$\theta$ derivatives of $W$ become dominant in equations (\ref{eqdui}) and
(\ref{eqduphi}), which then imply
\begin{equation}
\delta u_r\sim i{\sigma_0^2\over\sigma_0^2-\kappa_0^2}
{\partial W\over\partial r},\,\,\,\,\, \delta u_{\theta}\sim i{\partial W\over
\partial\theta},\,\,\,\,\,{\rm and}\,\,\,\,\,\delta u_\phi\sim{{\cal E}_0^2
\sigma_0\over\sigma_0^2-\kappa_0^2}g_0^{rr}\left({\partial\ell\over\partial r}
\right)_0{\partial W\over\partial r}.
\label{eqdus}
\end{equation}
Substituting into the linearized continuity equation, transforming coordinates
from $r$ and $\theta$ to the local, scaled coordinates $\bar{x}$ and
$\bar{y}$, and taking the slender torus limit $\beta\rightarrow0$, we
finally obtain
\begin{equation}
{\bar{\sigma}_0^2\over\bar{\sigma}_0^2-\bar{\kappa}_0^2}{\partial\over
\partial\bar{x}}\left(f^n{\partial W\over\partial\bar{x}}\right)+
{\partial\over\partial\bar{y}}\left(f^n{\partial W\over\partial\bar{y}}\right)
+2n\bar{\sigma}_0^2f^{n-1}W=0,
\label{eqwpde}
\end{equation}
where $\bar{\sigma}_0\equiv\sigma_0/\Omega_0\equiv(\omega-m\Omega_0)/\Omega_0$.

This partial differential equation must be solved together with the boundary
condition that the Lagrangian perturbation in pressure at the unperturbed
surface vanish:
\begin{equation}
\Delta p|_{f=0}=(\delta p+\xi^{\alpha}\nabla_\alpha p)|_{f=0}=0,
\end{equation}
where $\xi^\alpha$ is the Lagrangian displacement vector.
Using the gauge choice $\xi^\alpha u_\alpha=0$ \citep{ips92}, the
$r$ and $\theta$ components are given by
\begin{equation}
{\bxi}={i\over A\sigma}\delta{\bf u}.
\end{equation}
In the slender torus limit, the boundary condition may then be written as
\begin{equation}
\left\{f^n\left[{\bar{\sigma}_0\over\bar{\sigma}_0^2-\bar{\kappa}_0^2}
{\partial f\over\partial\bar{x}}{\partial W\over\partial\bar{x}}+
{1\over\bar{\sigma}_0}{\partial f\over\partial\bar{y}}
{\partial W\over\partial\bar{y}}+2\bar{\sigma}_0W\right]\right\}_{f=0}=0.
\label{eqbc}
\end{equation}

\section{General Remarks}

Equations (\ref{eqf}), (\ref{eqwpde}), and (\ref{eqbc}) describe a global
eigenvalue problem for modes of the relativistic, polytropic slender torus.
It is important to note that they are, necessarily, {\it identical} in
mathematical form
to the corresponding equations for a polytropic slender torus in a Newtonian
gravitational potential with axisymmetry and reflection symmetry.  All the
interesting effects boil down to three characteristic frequencies:  the
radial epicyclic frequency $\bar{\omega}_r$, the vertical epicyclic
frequency $\bar{\omega}_\theta$, and the characteristic frequency of
inertial modes at the pressure maximum $\bar{\kappa}_0$.

The slender torus always admits a trivial zero velocity, zero corotating
frequency mode, with $W$ being independent of $\bar{x}$ and $\bar{y}$, and
$\bar{\sigma}_0^2=0$.  In the constant specific angular momentum case, this
is the mode which corresponds to the principal mode of the Papaloizou-Pringle
instability when the torus has finite thickness \citep{bla85}.

As shown by \citet{abr05}, the slender torus (even in the baroclinic
case) also always admits incompressible modes corresponding to global
oscillations of the entire torus at the epicyclic frequencies of the
external spacetime:
\begin{equation}
W_r=\bar{x},\,\,\,\,\bar{\sigma}_0^2=\bar{\omega}_r^2
\end{equation}
and
\begin{equation}
W_\theta=\bar{y},\,\,\,\,\bar{\sigma}_0^2=\bar{\omega}_\theta^2.
\end{equation}
The fluid velocity for these modes is spatially constant over a torus
cross-section.

Broadly speaking, the remaining modes are of three general
types: surface gravity waves, 
internal inertial waves, and acoustic waves.  The last two classes
are easily revealed in a WKB analysis of equation (\ref{eqwpde}).
Assuming an $\bar{x}-\bar{y}$ dependence proportional to
$\exp\{i[\int \bar{k}_x(\bar{x})d\bar{x}+\int \bar{k}_y(\bar{y})d\bar{y}]\}$,
and taking the high wavenumber limit
$\bar{k}\equiv(\bar{k}_x^2+\bar{k}_y^2)^{1/2}\rightarrow\infty$,
we obtain the local dispersion relation
\begin{equation}
\bar{\sigma}_0^4-(\bar{k}^2\bar{c}_s^2+\bar{\kappa}_0^2)\sigma_0^2+
\bar{k}_y^2\bar{c}_s^2\bar{\kappa}_0^2=0,
\label{eqwkbdisp}
\end{equation}
where
\begin{equation}
\bar{c}_s\equiv{c_s\over\beta A_0 r_0\Omega_0}=
c_s\left({\rho_0\over2(n+1)p_0}\right)^{1/2}
\end{equation}
is a scaled local sound speed.  This dispersion relation immediately gives
two solutions describing the high wavenumber behavior for acoustic and
inertial waves:
\begin{equation}
\bar{\sigma}_0^2=\bar{k}^2\bar{c}_s^2\,\,\,\,\,{\rm and}\,\,\,\,\,
\bar{\sigma}_0^2={\bar{k}_y^2\bar{\kappa}_0^2\over\bar{k}^2},
\end{equation}
respectively.  This justifies our claim that $\bark$ is the characteristic
frequency of inertial oscillations in the fluid.

According to the equilibrium equation (\ref{eqf}), surfaces of constant
density and pressure have perfectly elliptical cross-section in
$(\bar{x},\bar{y})$ coordinates.  The slender torus equilibrium is therefore
perfectly reflection-symmetric about $\bar{x}=0$ and $\bar{y}=0$.  All the
slender torus modes can therefore be assigned a definite parity with respect
to each of these reflections, and we note these parities throughout the
rest of the paper.  In general, thicker tori remain reflection symmetric
about the equatorial plane, and so modes will continue to have a definite
$y-$parity.  A well-defined $x-$parity only exists in the slender torus
limit, however.

In addition to the parities, we will also label modes with non-negative
integer quantum numbers $j$ and $k$ which are derived from the extreme limits
of a constant specific angular momentum slender torus in a Newtonian point
mass potential and a general slender torus with an incompressible equation
of state $n=0$.  These limits are discussed in sections 4.1 and 4.3 below.
Physically, $j>0$ modes have an acoustic character, and have frequencies which
diverge in the incompressible limit.  Increasing $j$ and $k$ corresponds
to shorter spatial scales of variation in the mode eigenfunction on a
torus cross-section.

The tori we are analyzing in this paper are violently unstable
\citep{pap84}.  This instability is not captured by our
equations, because its growth rate happens to vanish in the slender torus
limit.  However, for tori of finite thickness, the corotation mode becomes
the principal mode of the Papaloizou-Pringle instability \citep{bla85,gol86}.
In addition, the slender torus limit loses the
corotation and Lindblad resonances normally associated with
nonaxisymmetric spiral waves in discs.

\section{Complete Solutions in Special Cases}

There are four special cases where a complete solution to the mode spectrum
of the slender torus can be derived analytically.

\subsection{Constant Specific Angular Momentum in a Newtonian $1/r$ Potential}

In a Newtonian $1/r$ potential, the two epicyclic frequencies
are both degenerate with the orbital angular velocity:
$\bar{\omega}_r=\bar{\omega}_\theta=1$.  If in addition the
torus has constant specific angular momentum, then $\bar{\kappa}_0^2=0$ and
equation (\ref{eqf}) implies that isobaric surfaces have circular cross-section
in the variables $\bar{x}$ and $\bar{y}$.  By changing variables to
$(\eta,\theta)$ given by
\begin{equation}
\bar{x}=\eta\cos\theta\,\,\,\,\,{\rm and}\,\,\,\,\,\bar{y}=\eta\sin\theta,
\end{equation}
Blaes (1985) showed that the partial differential equation (\ref{eqwpde})
becomes separable and solved the complete eigenvalue problem.
The eigenfunctions and eigenfrequencies may be conveniently written as
\begin{equation}
W_{jk}=C_{jk}\eta^{k}G_j(k+n,k+1,\eta^2)\left\{\matrix{\cos(k\theta)\cr
                                                       \rm{or}\cr
                                                       \sin(k\theta)}\right\}
\label{wjk}
\end{equation}
and
\begin{equation}
\bar{\sigma}_0^2={1\over n}(2j^2+2jn+2jk+nk)
\label{nujk}
\end{equation}
Here $C_{jk}$ is a normalization constant, $j$ and $k$ are non-negative
integers, and $G_j$ is a $j$-th order Jacobi polynomial
\citep{abrst72}.

The zero frequency, $j=k=0$ mode is the corotation mode.
The $j=0$, $k=1$ modes correspond
to the global radial and vertical epicyclic oscillations of the entire torus,
which in this case have frequencies that are degenerate with the orbital
angular velocity.  Other modes with $j=0$ have frequencies which are
independent of the polytropic equation of state, and represent surface
gravity waves.  Modes with $k=0$ and $j\ne0$ are compressible breathing
modes where the torus cross-section expands and contracts.  All
other modes have a mixed acoustic/surface gravity wave character.

\subsection{Keplerian Limit}

A specific angular momentum distribution which is at the opposite extreme to
that just considered is a ``Keplerian'' distribution, where every fluid element
in the slender torus has the same specific angular momentum as that of a
circular geodesic in the equatorial plane.  In this case the characteristic
frequency of inertial mode oscillations $\bark$ is equal to the radial
epicyclic frequency $\baromr$.  The eigenvalue problem in this limit becomes
very simple even in general axisymmetric spacetimes with reflection symmetry.

In the Keplerian limit, the equilibrium function $f=1-\baromt^2\bar{y}^2$
only varies with height $\bar{y}$.  One set of modes then has $W$ equal to
some linear function of $\bar{x}$ times some function of height $\bar{y}$.
For them, equation (\ref {eqwpde}) becomes 
\begin{equation}
{d\over d\bar{y}}\left(f^n{dW\over d\bar{y}}\right)
+2n\bar{\sigma}_0^2f^{n-1}W=0.
\end{equation}
The eigenfunctions are therefore simply Gegenbauer polynomials in $\bar{y}$
\citep{abrst72},
\begin{equation}
W=C^{(n-1/2)}_{n_y}(\baromt\bar{y})\,\,\,\,\,{\rm or}\,\,\,\,\,\bar{x}
C^{(n-1/2)}_{n_y}(\baromt\bar{y}),
\end{equation}
where $n_y$ is a non-negative integer which is equal to the number of vertical
nodes.  The corresponding eigenfrequencies are
\begin{equation}
\bar{\sigma}_0^2=\bar{\omega}_\theta^2\left[{n_y(n_y+2n-1)\over2n}\right].
\label{vertfreq}
\end{equation}
The lowest order, $n_y=0$ mode is the corotation mode, and $n_y=1$ gives the
vertical epicyclic mode.  Higher values of $n_y$ represent standing, vertical
acoustic waves.  (The eigenfrequencies for the $n_y=1$, 2, and 3 vertical
oscillations have also been derived by \citealt{kat05} within Newtonian
theory.)  The choice of $1$ or $\bar{x}$ for the radial dependence
indicates that these modes are at least doubly degenerate, and in fact we
show numerically in section 5 below that there are many degenerate modes
whose frequencies are given by equation (\ref{vertfreq}) at every $n_y$.
In addition, we find that all other modes, apart from the trivial corotation
mode, have frequencies which approach the radial epicyclic frequency.  Thus
the Keplerian limit of the slender torus consists of many degenerate modes
at the two epicyclic frequencies and frequencies corresponding to vertical
sound waves, plus the corotation mode.

\subsection{The Incompressible Torus}

In the limit $n\rightarrow 0$, the torus is incompressible and admits only
inertial and surface gravity waves (the latter being the $j=0$ modes discussed
in section 4.1 above).  These modes can be explicitly calculated for
incompressible tori in
the general spacetime.  In the incompressible limit, equations (\ref{eqwpde})
and (\ref{eqbc}) become
\begin{equation}
{\bar{\sigma}_0^2\over\bar{\sigma}_0^2-\bar{\kappa}_0^2}{\partial^2W\over
\partial\bar{x}^2}+{\partial^2W\over\partial\bar{y}^2}=0
\label{eqwpden0}
\end{equation}
and
\begin{equation}
\left[{\bar{\sigma}_0\over\bar{\sigma}_0^2-\bar{\kappa}_0^2}
{\partial f\over\partial\bar{x}}{\partial W\over\partial\bar{x}}+
{1\over\bar{\sigma}_0}{\partial f\over\partial\bar{y}}
{\partial W\over\partial\bar{y}}+2\bar{\sigma}_0W\right]_{f=0}=0.
\label{eqbcn0}
\end{equation}
For modes with $\bar{\sigma}_0^2>\bark^2$, it is convenient to change
coordinates from $\bar{x}$ to
\begin{equation}
\chi\equiv\left({\bar{\sigma}_0^2-\bar{\kappa}_0^2\over\bar{\sigma}_0^2}
\right)^{1/2}\bar{x},
\label{eqchidef}
\end{equation}
so that equation (\ref{eqwpden0}) becomes Laplace's equation:
\begin{equation}
{\partial^2W\over\partial\chi^2}+{\partial^2W\over\partial\bar{y}^2}=0.
\label{eqwlaplace}
\end{equation}
The boundary condition (\ref{eqbcn0}) also simplifies:
\begin{equation}
\left[{1\over\bar{\sigma}_0}
{\partial f\over\partial\chi}{\partial W\over\partial\chi}+
{1\over\bar{\sigma}_0}{\partial f\over\partial\bar{y}}
{\partial W\over\partial\bar{y}}+2\bar{\sigma}_0W\right]_{f=0}=0.
\label{eqbcchi}
\end{equation}
In terms of $\chi$, the equilibrium function $f$ now becomes, from equation
(\ref{eqf}),
\begin{equation}
f=1-{\cal R}^2\chi^2-\baromt^2\bar{y}^2,
\label{eqfchi}
\end{equation}
where
\begin{equation}
{\cal R}\equiv\left[{\bar{\sigma}_0^2(\bar{\omega}_r^2-\bar{\kappa}_0^2)
\over\bar{\sigma_0}^2-\bar{\kappa}_0^2}\right]^{1/2}.
\label{eqcalr}
\end{equation}

It is well-known that Laplace's equation (\ref{eqwlaplace}) separates in
confocal elliptical coordinates $(u,v)$, defined by
\begin{equation}
\chi=a\cosh u\cos v
\end{equation}
and
\begin{equation}
\bar{y}=a\sinh u\sin v,
\end{equation}
where $u$ is a non-negative real number and $0\le v<2\pi$.  We choose
\begin{equation}
a=\left({1\over{\cal R}^2}-{1\over\bar{\omega}_\theta^2}\right)^{1/2},
\end{equation}
in order that the surface $f=0$ of the torus corresponds to one of the
elliptical coordinate surfaces.  This has a fixed value of $u$ given by
\begin{equation}
u=\tanh^{-1}\left({{\cal R}\over\baromt}\right)=
{1\over2}\ln\left({\baromt+{\cal R}\over\baromt-{\cal R}}\right).
\end{equation}
The resulting eigenfunctions are then
\begin{equation}
W_k^+=C_k^+\cosh(ku)\cos(kv)\,\,\,\,{\rm and}\,\,\,\,W_k^-=C_k^-\sinh(ku)
\sin(kv),
\label{eqwinc1}
\end{equation}
where the $C$'s are again normalization constants and $k$ is again a
non-negative integer.
The corresponding eigenfrequencies are given by
\begin{equation}
\bar{\sigma}_0^2=k{\cal R}\bar{\omega}_\theta\left[{(\bar{\omega}_\theta+
{\cal R})^k-(\bar{\omega}_\theta-{\cal R})^k\over(\bar{\omega}_\theta+
{\cal R})^k+(\bar{\omega}_\theta-{\cal R})^k}\right]^{\pm1}.
\label{eqsiginc}
\end{equation}

Our derivation of the eigenfunctions (\ref{eqwinc1}) and eigenfrequencies
(\ref{eqsiginc}) assumed that $\bars^2>\bark^2$.  However, equation
(\ref{eqsiginc}) generally admits solutions with $\bars^2<\bark^2$ as well.
The reader may easily repeat the derivation for this case, and verify that
the resulting formula for the eigenfrequencies still turns out to be
equation (\ref{eqsiginc}).

Because ${\cal R}$ itself depends on $\bars$, equation (\ref{eqsiginc})
represents a polynomial equation for the eigenfrequency which must be solved
for every $k$.  The universal corotation mode is given by $k=0$, while $k=1$
gives the two epicyclic modes.  For modes with $k>1$, $\bars^2>\bark^2$
represent surface gravity waves, while $\bars^2<\bark^2$ represent internal
inertial waves.
For reference, we give explicit expressions of the eigenfrequencies and
eigenfunctions of some of the lowest order incompressible modes for general
angular momentum distributions in Table 1.\footnote{The reader may be surprised
that modes with the same eigenfunction can have different eigenfrequencies,
as is the case for the fourth row of Table 1.  However, the relationships
between $W$ and the Eulerian perturbations $\delta p$, eq. (\ref{eqwdef}),
and $\delta u_\alpha$, eq. (\ref{eqdus}), depend on frequency, so the fluid
perturbations are not the same.}

\begin{table*}
 \centering
 \begin{minipage}{140mm}
 \caption{Lowest Order Modes of the Incompressible Slender Torus.}
 \begin{tabular}{@{}ccccc@{}}
 \hline
 $x$-parity & $y$-parity & $k$ & $\bar{\sigma}_0^2$ & Eigenfunction\\
 \hline
 $+$ & $+$ & 0 & 0 & $1$ \\
 $-$ & $+$ & 1 & $\baromr^2$ & $\bar{x}$ \\
 $+$ & $-$ & 1 & $\baromt^2$ & $\bar{y}$ \\
 $-$ & $-$ & 2 & $\{\baromr^2+\baromt^2\pm
                 [(\baromr^2+\baromt^2)^2-
                 4\bar{\kappa}_0^2\baromt^2]^{1/2}\}/2$
               & $\bar{x}\bar{y}$ \\
 $+$ & $+$ & 2 & $\baromt^2(4\baromr^2-3\bar{\kappa}_0^2)/
                     (\baromt^2+\baromr^2-\bar{\kappa}_0^2)$
               & $1-{(4\baromt^2-\bark^2)(\baromr^2-\bark^2)\bar{x}^2-
                 \baromt^2(4\baromr^2-3\bark^2)\bar{y}^2\over
                 2\baromt^2-2\baromr^2+\bark^2}$ \\
 \hline
 \end{tabular}
 \end{minipage}
\end{table*}

The modes become much simpler in the constant specific angular momentum case
$\bar{\kappa}_0=0$.  Then ${\cal R}=\bar{\omega}_r$, and so
\begin{equation}
\bar{\sigma}_0^2=k\bar{\omega}_r\bar{\omega}_\theta\left[{(\bar{\omega}_\theta+
\bar{\omega}_r)^k-(\bar{\omega}_\theta-\bar{\omega}_r)^k
\over(\bar{\omega}_\theta+\bar{\omega}_r)^k+(\bar{\omega}_\theta-
\bar{\omega}_r)^k}\right]^{\pm1}.
\label{eqsiginckappa0}
\end{equation}
There are no internal inertial modes in this case, and these are all surface
gravity waves.  In the limit of a Newtonian $1/r$ potential,
$\bar{\omega}_r=\bar{\omega}_\theta=1$, and equation (\ref{eqsiginckappa0})
gives the $j=0$ surface gravity waves of equation (\ref{nujk}).

In the Keplerian limit, all the incompressible modes become degenerate with
the radial and vertical epicyclic modes.  The inertial modes all have
frequencies approaching the radial epicyclic frequency, while the surface
gravity waves approach either the radial or the vertical epicyclic frequencies.

\subsection{Neglecting the Vertical Structure}

In an effort to understand the axisymmetric modes seen in numerical studies
of tori, \citet{rez03b} and \citet{mon04} numerically calculated the
eigenfunctions and eigenfrequencies of vertically integrated tori.
This begs the question - what are the slender torus modes if we ignore the
vertical direction?\footnote{As \citet{rez03b} point out, one must be wary when
comparing polytropic, three-dimensional tori with their two-dimensional,
vertically integrated representations.  In general, the latter do not
satisfy a polytropic relationship between vertically integrated pressure and
density.  In the slender torus limit, however, there is an exact mapping
between polytropic equations of state in the equilibrium structure, with
$n_{\rm 2D}=n_{\rm 3D}+1/2$
\citep{gol86}.  Ignoring the vertical direction is therefore equivalent to
adopting the polytropic index $n_{\rm 2D}$.  Note that \citet{rez03b} and
\citet{mon04} choose $n_{\rm 2D}=3$ for their numerical results.} The problem
is easy to solve in general, and is mathematically similar to the Keplerian
limit vertical acoustic waves discussed in section 4.2 above.  The
eigenfunctions are again Gegenbauer polynomials:
\begin{equation}
W=C^{(n-1/2)}_{n_x}[(\baromr^2-\bark^2)^{1/2}\bar{x}],
\end{equation}
where $n_x$ is a non-negative integer equal to the number of radial nodes.
The corresponding eigenfrequencies are
\begin{equation}
\bar{\sigma}_0^2=\bar{\kappa}_0^2+(\bar{\omega}_r^2-\bar{\kappa}_0^2)
\left[{n_x(n_x+2n-1)\over2n}\right].
\end{equation}
Note that $n_x=1$ gives the radial epicyclic mode.  \citet{rez03b} and
\citet{mon04} labeled this mode as an $f$-mode, and higher order modes as
$o_1$, $o_2$, etc., i.e. $o_{n_x-1}$.  As they point out, this latter set of
modes are physically acoustic (or $p$-)modes, at least in this two dimensional
situation.  However, as we show below in the next section, some of these
modes are actually related to surface gravity waves in three dimensions,
at least in the slender torus limit.  This is true in particular of the $o_1$
mode, which with the radial epicyclic mode they propose might be responsible
for the observed $3:2$ commensurability in high frequency QPO's in black
hole X-ray binaries \citep{rez03a}.

The ratio in frequencies of the two-dimensional $p$-modes to the radial
epicyclic mode is given by
\begin{equation}
{\bars\over\baromr}=\pm\baromr\left[n_x\left(1+{n_x-1\over2n}\right)-(n_x-1)
\left(1+{n_x\over2n}\right){\bark^2\over\baromr^2}\right]^{1/2}.
\label{eqfrat2d}
\end{equation}
For constant specific angular momentum and $n=3$, the sequence of frequency
ratios is approximately 1.528, 2, 2.449, ..., in good agreement with the
$(3/2,4/2,5/2)$ sequence that \citet{rez03b} claimed was an approximate
description of their low order numerical eigenfrequencies.  The number 1.528
is also in excellent agreement with the $o_1/f$ ratio found numerically by
\citet{mon04} (see their Figure 6) to be generic for all slender, vertically
integrated, relativistic tori with constant specific angular momentum.  They
also found that the frequency ratio dropped for non-constant specific
angular momentum (see their Figure 8), at least for slender tori, consistent
with the fact that the frequency ratios in equation (\ref{eqfrat2d}) all
approach unity in the Keplerian limit.
We discuss the frequency ratios of the lowest order modes in {\it three}
dimensional tori in the next section.

\section{The Lowest Order Modes of the General Problem}

As we noted in section 3, the zero corotation frequency and epicyclic modes
are always present for the general torus configuration.
One can find additional modes by following the suspicion that a Taylor
expansion solution for $W$ about the pressure maximum will generally diverge as
the surface $f=0$ is approached because of the structure of the partial
differential equation (\ref{eqwpde}).  A solution which is finite there, and
therefore able to satisfy the surface boundary condition (\ref{eqbc}), will
therefore probably involve a truncation of the Taylor expansion, i.e. involve
only finite polynomials in $\bar{x}$ and $\bar{y}$.  This is in fact true of
all the special case modes considered above.

Guided by the fact that the modes must be even or odd, we try an eigenfunction
of the form
\begin{equation}
W=a\bar{x}\bar{y},
\end{equation}
where $a$ is some constant. This function is odd in both $\bar{x}$ and
$\bar{y}$, and clearly satisfies the boundary condition (\ref{eqbc}).  It gives
in fact two modes, with eigenfrequencies determined by the partial differential
equation (\ref{eqwpde}):
\begin{equation}
\bar{\sigma}_0^2={1\over2}\left\{\baromr^2+\baromt^2\pm\left[(\baromr^2+
\baromt^2)^2-4\bark^2\baromt^2\right]^{1/2}\right\}.
\end{equation}
These modes are identical to the odd-odd $k=2$ modes of the incompressible
torus.  The positive square root gives a purely incompressible surface gravity
wave whose poloidal velocity field from equation (\ref{eqdus}) is
reminiscent of an X-mode polarization
gravitational wave.  The negative square root gives a purely incompressible
inertial mode whose poloidal velocity field represents a circulation
around the pressure maximum.

We can go further.  Try an eigenfunction of the form
\begin{equation}
W=a+b\bar{x}^2+c\bar{y}^2,
\end{equation}
which is even in both $\bar{x}$ and $\bar{y}$.  Here $a$, $b$, and $c$ are
constants, one of which is arbitrary.  Once again, the boundary condition is
trivially satisfied.  Substituting this function into equation (\ref{eqwpde}),
and equating coefficients of $1$, $\bar{x}^2$ and $\bar{y}^2$, we end up with
three linear homogeneous equations for $a$, $b$, and $c$.  Setting the
determinant of coefficients to zero, we obtain the eigenfrequencies of the
zero corotation frequency mode (for which $b$=$c$=0), as well as
\begin{equation}
\bar{\sigma}_0^2={1\over2n}\left\{(2n+1)(\baromt^2+\baromr^2)-(n+1)\bark^2
\pm\left[\biggl((2n+1)(\baromt^2-\baromr^2)+(n+1)\bark^2\biggr)^2+4\baromt^2
(\baromr^2-\bark^2)\right]^{1/2}\right\}
\label{eqplusmodefreq}
\end{equation}
The corresponding values of $b$ and $c$ in terms of $a$ are straightforward
to calculate, and can be found in Table 3.

The frequency
corresponding to the upper sign in equation (\ref{eqplusmodefreq}) blows
up in the incompressible $n\rightarrow0$ limit, revealing that it is
in general an acoustic mode.  It reduces in fact
to the acoustic $j=1$, $k=0$ mode in the limit of a constant specific
angular momentum torus in a Newtonian $1/r$ potential, which we discussed
in section 4.1.
The velocity field in that case is a breathing mode - the torus cross-section
expands and contracts.  In the Keplerian limit of section 4.2, the mode
frequency becomes that of the $n_y=2$ vertical acoustic wave.

The frequency corresponding to the lower sign in equation
(\ref{eqplusmodefreq}) reduces to the even-even $k=2$ mode of Table 1
in the incompressible, $n\rightarrow0$ limit.  This is a gravity wave,
and the velocity field in this limit is reminiscent of a $+$-mode
polarization gravitational wave.

\begin{table*}
 \centering
 \begin{minipage}{140mm}
 \caption{Eigenfrequencies of the Lowest Order Modes of the General
Polytropic Slender Torus.}
 \begin{tabular}{@{}ccccc@{}}
 \hline
$x$-parity & $y$-parity & $j$ & $k$ &
$\bar{\sigma}_0^2$ \\
\hline
$+$ & $+$ & 0 & 0 & 0 \\
$-$ & $+$ & 0 & 1 & $\baromr^2$ \\
$+$ & $-$ & 0 & 1 & $\baromt^2$ \\
$-$ & $-$ & 0 & 2 & $\{\baromr^2+\baromt^2\pm
                 [(\baromr^2+\baromt^2)^2-
                 4\bar{\kappa}_0^2\baromt^2]^{1/2}\}/2$ \\
$+$ & $+$ & 0 & 2 & $\{(2n+1)(\baromt^2+\baromr^2)-(n+1)
                     \bar{\kappa}_0^2-[((2n+1)(\baromt^2-
                     \baromr^2)+(n+1)\bar{\kappa}_0^2)^2+
                     4\baromt^2(\baromr^2-
                     \bar{\kappa}_0^2)]^{1/2}\}/(2n)$ \\
$+$ & $+$ & 1 & 0 & $\{(2n+1)(\baromt^2+\baromr^2)-(n+1)
                     \bar{\kappa}_0^2+[((2n+1)(\baromt^2-
                     \baromr^2)+(n+1)\bar{\kappa}_0^2)^2+
                     4\baromt^2(\baromr^2-
                     \bar{\kappa}_0^2)]^{1/2}\}/(2n)$ \\
\hline
 \end{tabular}
 \end{minipage}
\end{table*}

\begin{table*}
 \centering
 \begin{minipage}{140mm}
 \caption{Eigenfunctions of the Lowest Order Modes of the General
Polytropic Slender Torus.}
 \begin{tabular}{@{}cccc@{}}
 \hline
$x$-parity & $y$-parity & $(j,k)$ & Eigenfunction\\
\hline
$+$ & $+$ & (0,0) & $1$ \\
$-$ & $+$ & (0,1) & $\bar{x}$ \\
$+$ & $-$ & (0,1) & $\bar{y}$ \\
$-$ & $-$ & (0,2) & $\bar{x}\bar{y}$ \\
$+$ & $+$ & (0,2) or (1,0) & $1+{[2n\bars^2-4(n+1)\baromt^2+\bark^2]
(\baromr^2-\bark^2)\bar{x}^2-[2n\bars^2-4(n+1)\baromr^2+(2n+3)\bark^2]
\baromt^2\bar{y}^2\over2\baromt^2-2\baromr^2+\bark^2}$ \\
\hline
 \end{tabular}
 \end{minipage}
\end{table*}

\begin{figure}
\includegraphics[width=180mm]{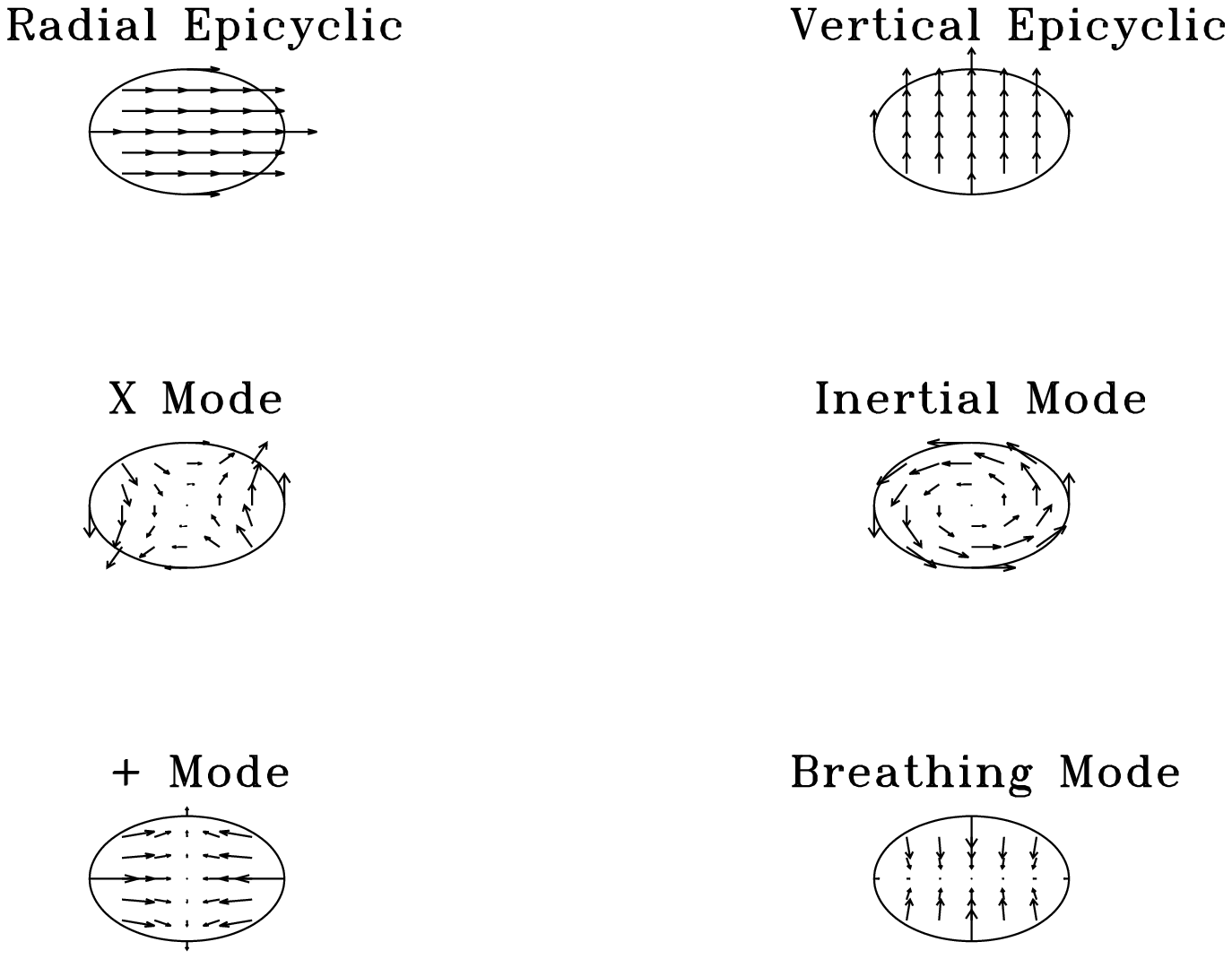}
\caption{Poloidal velocity fields $(\delta u_x,\delta u_y)$ of the lowest
order, nontrivial slender torus modes.  The torus was assumed to be orbiting
in a Kerr spacetime with $a/M=0.5$ and at a Boyer-Lindquist coordinate
radius of $r_0=10M$.  The slope of the internal specific angular momentum
distribution was assumed to give $\bark/\baromr=0.5$ and the polytropic index
was $n=3$.
\label{modevelocity}}
\end{figure}

We summarize the frequencies and eigenfunctions of all the lowest order modes
of the general slender torus in Tables 2 and 3.  We also show the poloidal
velocity fields for these modes in Figure \ref{modevelocity} for a torus
with $\bark/\baromr=0.5$.  While we are still quite far from the Keplerian
limit, the velocity fields are already showing what happens in that case.
The X-mode frequency becomes degenerate with $\baromt$ in that limit, consisting
of opposite vertical oscillations on either radial side of the pressure
maximum.  The breathing mode becomes degenerate with the lowest order vertical
acoustic wave, as it's velocity field becomes largely vertical.  The inertial
and +-modes consist largely of radial motions in the Keplerian limit, and
become degenerate with $\baromr$.

It is clear that the procedure we used in this section can be extended to
calculate even higher order modes.  In the special case of
constant specific angular momentum tori, the dispersion relation remains
quadratic for modes of quite high order, and we summarize the frequencies and
eigenfunctions of these modes in Tables 4 and 5.

In general, however, higher order modes result in dispersion relations which
are polynomials of cubic and higher order, and should be solved numerically.
To do this, we substitute the power-series expansion
\begin{eqnarray}
W(\bar{x},\bar{y}) & = & \sum_{i,l=0}^\infty W_{il}\bar{x}^i
\bar{y}^l.
\label{eq:powerseries}
\end{eqnarray}
in equation (\ref{eqwpde}), and equate coefficients of $\bar{x}^i
\bar{y}^l$. We find the following set of algebraic equations
\begin{eqnarray}
&& \frac{\bar{\sigma}_0^2}{\bar{\sigma}_0^2-\bar{\kappa}_0^2} 
 \left[  (i+1)(i+2)W_{i+2,l}-(\baromr^2
  -\bar{\kappa}_0^2)i(i-1) W_{il} \right.
 \nonumber \\ &  -  & \left. 
  \baromt^2(i+1)(i+2)W_{i+2,l-2}  \right]
+ (l+1)(l+2)W_{i,l+2}
\nonumber \\ & & 
-(\baromr^2
  -\bar{\kappa}_0^2)(l+1)(l+2)W_{i-2,l+2}-\baromt^2l(l-1)W_{il}
\nonumber \\ & & 
- 2n 
\frac{\bar{\sigma}_0^2}{\bar{\sigma}_0^2-\bar{\kappa}_0^2}(\baromr^2
 -\bar{\kappa}_0^2) i W_{il}
- 2n \baromt^2 l W_{il} 
+ 2n \bar{\sigma}_0^2 W_{il} = 0,
\label{eq:matrixeqn}
\end{eqnarray}
which can be written as a matrix equation $A(\bar{\sigma}_0)W=0$,
defining the matrix $A$.
We solve equation (\ref{eq:matrixeqn}) with basis vectors for $i=i_1+2(I-1)$
and $l=l_1+2(L-1)$, where $I=1,...,I_{\rm max}$, $L=1,...,L_{\rm max}$.
Here $i_1=0$ for even $x$-parity and $1$ for odd $x$-parity, and
similarly for $l_1$.  Our numerical technique to solve
equation (\ref{eq:matrixeqn}) is to sweep through frequency space and look for
zeros of the determinant of the matrix $A(\bar{\sigma}_0)$.

\begin{figure}
 \includegraphics[width=160mm]{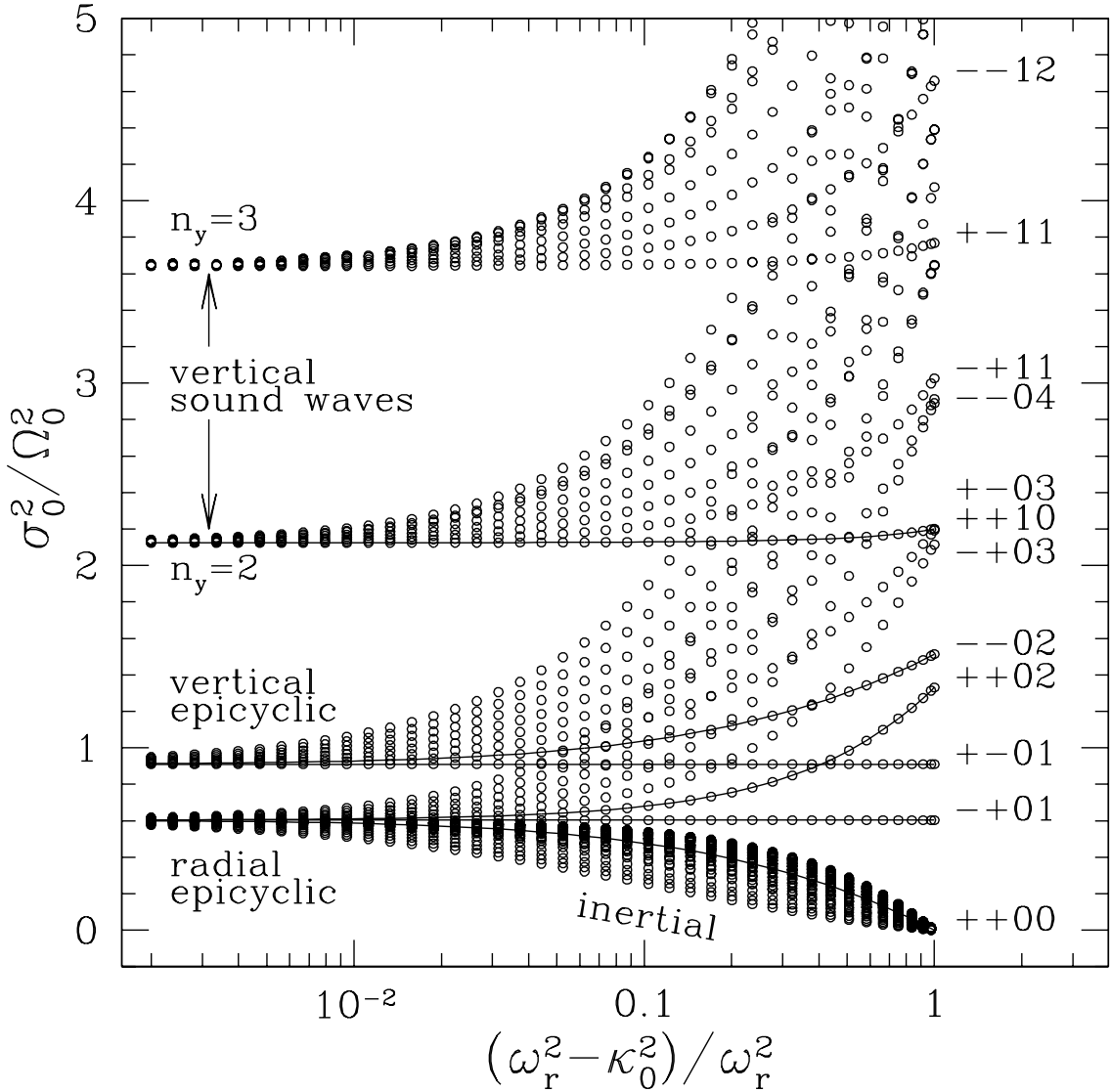}
 \caption{Slender torus eigenfrequencies as a function of
  $(\omega_r^2-\kappa_0^2)/\omega_r^2$, which ranges from the
  Keplerian limit (left) to the constant specific angular momentum
  limit (right). Circles show the numerical solution to equation
  (\ref{eqwpde}), assuming four basis functions for $\bar{x}$ and $\bar{y}$.
  Lines show the exact analytic solutions in Table 2.
  Vertical sound waves in the Keplerian limit are labeled by the
  number of vertical nodes $n_y=2,3$ (see equation \ref{vertfreq}).
  Labels on the right hand side are for modes in a
  constant specific angular momentum torus (Table 4).
  The black hole was taken to have $a/M=0.9$ and the torus
  centre was at $r=10M$, giving $\baromr^2=0.604$ and
  $\baromt^2=0.910$. The polytropic index was $n=3$.  The corotation
  mode ($\bar{\sigma}_0^2=0$) is not shown, but is also present
  for all specific angular momentum distributions.
\label{fig:sigmavskappa}}
\end{figure}

Figure \ref{fig:sigmavskappa} shows eigenfrequencies found by solving
equation (\ref{eq:matrixeqn}) compared to the various analytical
results. The solutions are greatly simplified in the Keplerian
limit, as many different branches converge together. The lowest clump
approaches the radial epicyclic mode at $\omega_r$. Inertial modes
approach $\omega_r$ from below in the Keplerian limit and go to
zero in the constant specific angular momentum limit. Surface gravity
waves approach $\omega_r$ from
above in the Keplerian limit. The next grouping is around the vertical
epicyclic mode at 
$\omega_\theta$. Surface gravity waves approach this frequency
from above. The next two groupings shown are sound waves with 2 and 3 vertical
nodes. The frequency increases as more horizontal nodes are added,
giving the ``fan'' away from the Keplerian limit. The lowest order
sound wave is nearly independent of $\kappa_0$, and higher order modes
are more sensitive.

\begin{table*}
 \centering
 \begin{minipage}{140mm}
 \caption{Eigenfrequencies of the Lowest Order Modes of the Constant
Specific Angular Momentum Slender Torus.}
 \begin{tabular}{@{}ccccc@{}}
 \hline
$x$-parity & $y$-parity & $j$ & $k$ & $\bar{\sigma}_0^2$ \\
 \hline
$+$ & $+$ & 0 & 0 & 0 \\
$-$ & $+$ & 0 & 1 & $\baromr^2$ \\
$+$ & $-$ & 0 & 1 & $\baromt^2$ \\
$-$ & $-$ & 0 & 2 & $\baromr^2+\baromt^2$ \\
$+$ & $+$ & 0 & 2 & $\{(2n+1)(\baromr^2+\baromt^2)-
                     [4n(n+1)(\baromt^2-\baromr^2)^2+
                     (\baromr^2+\baromt^2)^2]^{1/2}\}/(2n)$ \\
$+$ & $+$ & 1 & 0 & $\{(2n+1)(\baromr^2+\baromt^2)+
                     [4n(n+1)(\baromt^2-\baromr^2)^2+
                     (\baromr^2+\baromt^2)^2]^{1/2}\}/(2n)$ \\
$-$ & $+$ & 0 & 3 & $\{(4n+3)\baromr^2+(2n+1)\baromt^2-
                     [4n^2(\baromt^2-\baromr^2)^2+4n(\baromt^2-\baromr^2)
                     (\baromt^2-3\baromr^2)+(\baromt^2+3\baromr^2)^2]^{1/2}\}
                     /(2n)$ \\
$-$ & $+$ & 1 & 1 & $\{(4n+3)\baromr^2+(2n+1)\baromt^2+
                     [4n^2(\baromt^2-\baromr^2)^2+4n(\baromt^2-\baromr^2)
                     (\baromt^2-3\baromr^2)+(\baromt^2+3\baromr^2)^2]^{1/2}\}
                     /(2n)$ \\
$+$ & $-$ & 0 & 3 & $\{(4n+3)\baromt^2+(2n+1)\baromr^2-
                     [4n^2(\baromt^2-\baromr^2)^2+4n(\baromr^2-\baromt^2)
                     (\baromr^2-3\baromt^2)+(\baromr^2+3\baromt^2)^2]^{1/2}\}
                     /(2n)$ \\
$+$ & $-$ & 1 & 1 & $\{(4n+3)\baromt^2+(2n+1)\baromr^2+
                     [4n^2(\baromt^2-\baromr^2)^2+4n(\baromr^2-\baromt^2)
                     (\baromr^2-3\baromt^2)+(\baromr^2+3\baromt^2)^2]^{1/2}\}
                     /(2n)$ \\
$-$ & $-$ & 0 & 4 & $\{(4n+3)(\baromt^2+\baromr^2)-
                     [4n(n+3)(\baromt^2-\baromr^2)^2
                     +9(\baromt^2+\baromr^2)^2]^{1/2}\}
                     /(2n)$ \\
$-$ & $-$ & 1 & 2 & $\{(4n+3)(\baromt^2+\baromr^2)+
                     [4n(n+3)(\baromt^2-\baromr^2)^2
                     +9(\baromt^2+\baromr^2)^2]^{1/2}\}/(2n)$ \\
 \hline
 \end{tabular}
 \end{minipage}
\end{table*}

\begin{table*}
 \centering
 \begin{minipage}{140mm}
 \caption{Eigenfunctions of the Lowest Order Modes of the Constant
Specific Angular Momentum Slender Torus.}
 \begin{tabular}{@{}cccc@{}}
 \hline
$x$-parity & $y$-parity & $(j,k)$ & Eigenfunction\\
 \hline
$+$ & $+$ & (0,0) & $1$ \\
$-$ & $+$ & (0,1) & $\bar{x}$ \\
$+$ & $-$ & (0,1) & $\bar{y}$ \\
$-$ & $-$ & (0,2) & $\bar{x}\bar{y}$ \\
$+$ & $+$ & (0,2) or (1,0) & $1+{n\bar{\sigma}_0^2\over\baromt^2-\baromr^2}
(\baromr^2\bar{x}^2-\baromt^2\bar{y}^2)-{2(n+1)\baromt^2\baromr^2\over\baromt^2-
\baromr^2}(\bar{x}^2-\bar{y}^2)$ \\
$-$ & $+$ & (0,3) or (1,1) & $\bar{x}-{\baromt^2[(2n+1)\bar{\sigma}_0^2-3(2n+3)
\baromr^2]\over\bar{\sigma}_0^2-3\baromr^2}\left[\bar{x}\bar{y}^2+
{\baromr^2\over n\bar{\sigma}_0^2-3(n+1)\baromr^2}\bar{x}^3\right]$ \\
$+$ & $-$ & (0,3) or (1,1) & $\bar{y}-{\baromr^2[(2n+1)\bar{\sigma}_0^2-3(2n+3)
\baromt^2]\over\bar{\sigma}_0^2-3\baromt^2}\left[\bar{x}^2\bar{y}+{\baromt^2
\over n\bar{\sigma}_0^2-3(n+1)\baromt^2}\bar{y}^3\right]$ \\
$-$ & $-$ & (0,4) or (1,2) & $\bar{x}\bar{y}-{(2n+3)\baromr^2(\bar{\sigma}_0^2
-\baromr^2)-3(2n+5)\baromr^2\baromt^2\over3(\bar{\sigma}_0^2-\baromr^2-
3\baromt^2)}\bar{x}^3\bar{y}-{(2n+3)\baromt^2(\bar{\sigma}_0^2-\baromt^2)-
3(2n+5)\baromr^2\baromt^2\over3(\bar{\sigma}_0^2-\baromt^2-3\baromr^2)}
\bar{x}\bar{y}^3$ \\
 \hline
 \end{tabular}
 \end{minipage}
\end{table*}

\subsection{Mode Frequency Ratios}

Based on the behavior of its counterpart in
vertically integrated tori, \citet{rez03a} and \citet{zan05} identify the
``plus-mode'' surface gravity wave as a $p$-mode, which they label as the
$o_1$ mode.  For compressible tori, this characterization is not
entirely inaccurate.  While the radial compressions present
in this mode are always accompanied by vertical expansions in three
dimensions, these vertical expansions can be much smaller than the
radial motions if the torus is very compressible, as shown in Figure
\ref{modevelocity}.

\begin{figure}
 \includegraphics[width=160mm]{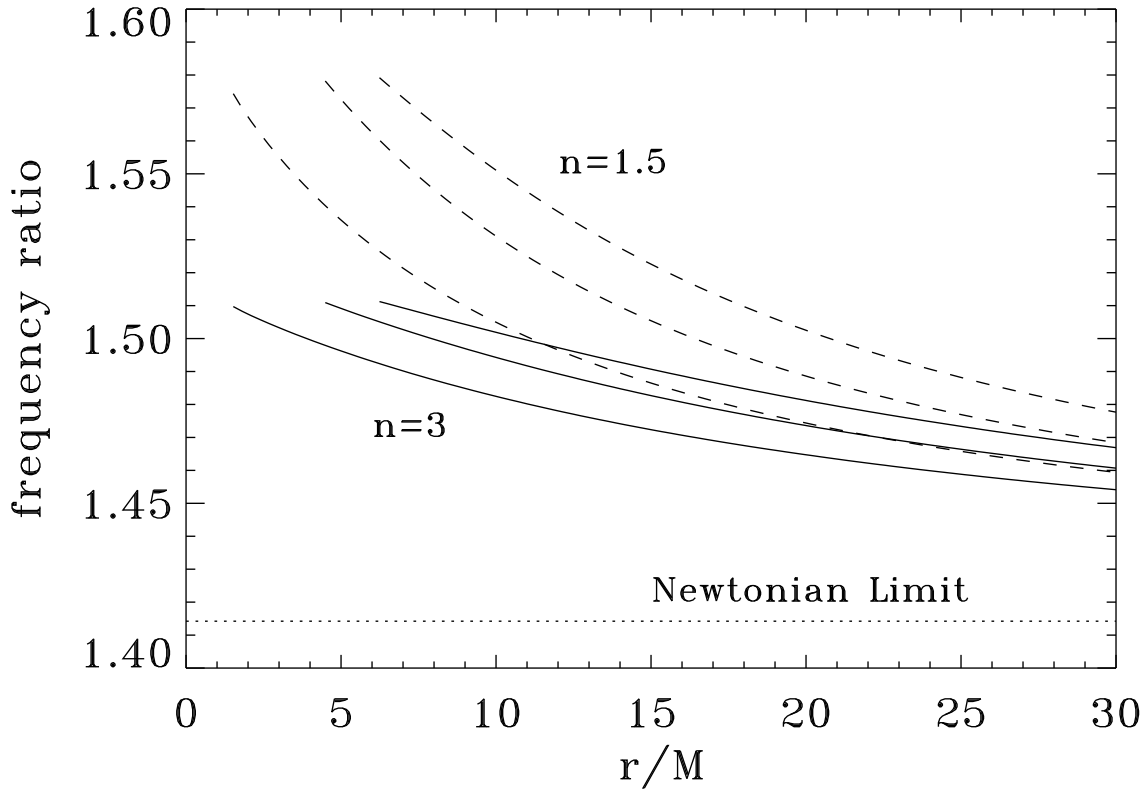}
 \caption{Ratio of the ``plus mode'' surface gravity wave $(+,+,j=0,k=2)$
frequency to the radial epicyclic mode frequency, for a slender constant
specific angular momentum $(\kappa_0^2=0$) torus in a Kerr spacetime at
different Boyer-Lindquist coordinate radii.  Solid curves are for a  
polytropic index $n=3$, while dashed curves are for $n=1.5$.  From top to
bottom, each of the three curves at fixed $n$ corresponds to  
spin parameters $a/M=0$, 0.5, and 0.998, respectively.  The Newtonian limit
of the frequency ratio, $2^{1/2}$, is indicated by the horizontal dotted line.
\label{figplusmode}}
\end{figure}

\begin{figure}
 \includegraphics[width=160mm]{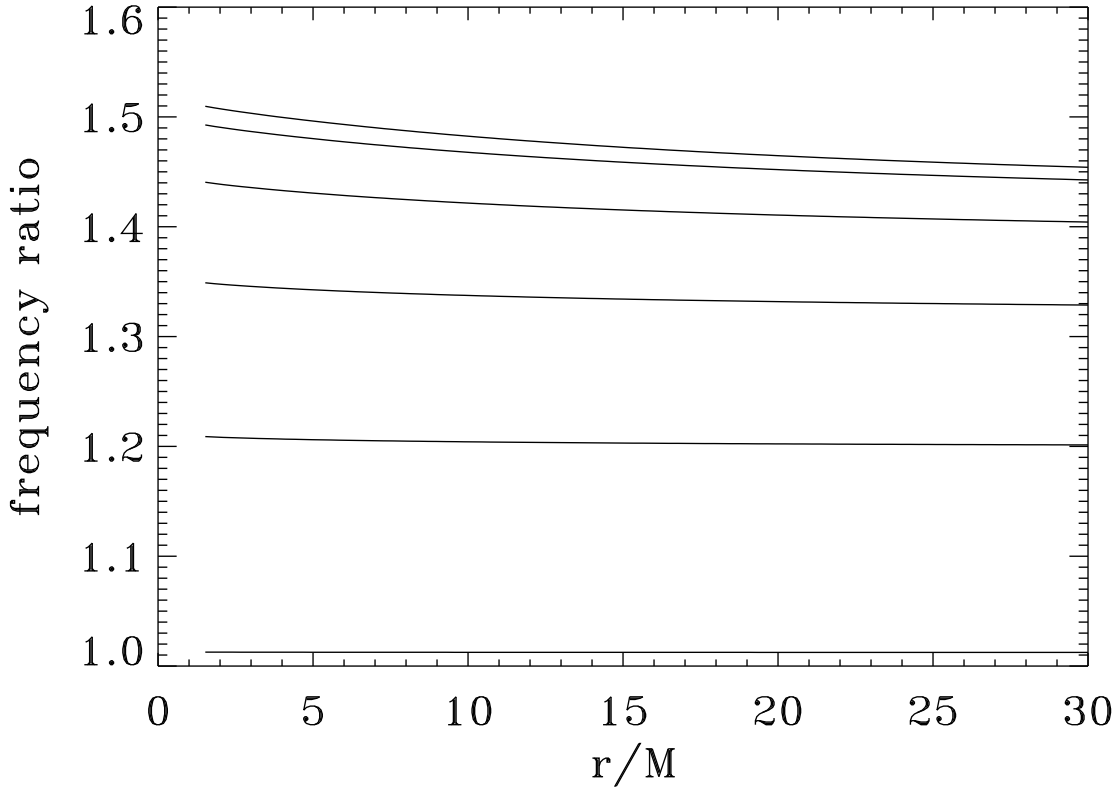}
 \caption{Ratio of the ``plus mode'' surface gravity wave $(+,+,j=0,k=2)$
frequency to the radial epicyclic mode frequency, for a slender $n=3$ torus
in an $a/M=0.998$ Kerr spacetime at different Boyer-Lindquist coordinate
radii and different values of $\kappa_0$.  From top to bottom, the curves
have $\kappa_0/\omega_r=0$, 0.2, 0.4, 0.6, 0.8, and 0.99. 
\label{figplusmoderkr}}
\end{figure}

It is this mode and the radial epicyclic mode
that \citet{rez03a} propose might be responsible for the 3:2 frequency ratio
observed in high frequency QPO's of black hole X-ray binaries.  Figure
\ref{figplusmode} depicts this frequency ratio computed from equation
(\ref{eqplusmodefreq}) for
slender, constant specific angular momentum tori of different polytropic
indices, orbiting Kerr black holes with various spin parameters, as a function
of the radius of the torus.  As \citet{rez03a}
found numerically in various specific cases, the frequency ratio is indeed
close to 1.5.  Relativity is essential here, as the ratio drops to $2^{1/2}$
in the Newtonian limit of large torus radii (see eq. \ref{nujk} with $j=0$
and $k=2$).  As shown in Figure \ref{figplusmoderkr}, the frequency ratio
also drops for non-constant specific angular momentum, approaching unity
in the Keplerian limit $\bark^2=\baromr^2$.

\begin{figure}
 \includegraphics[width=160mm]{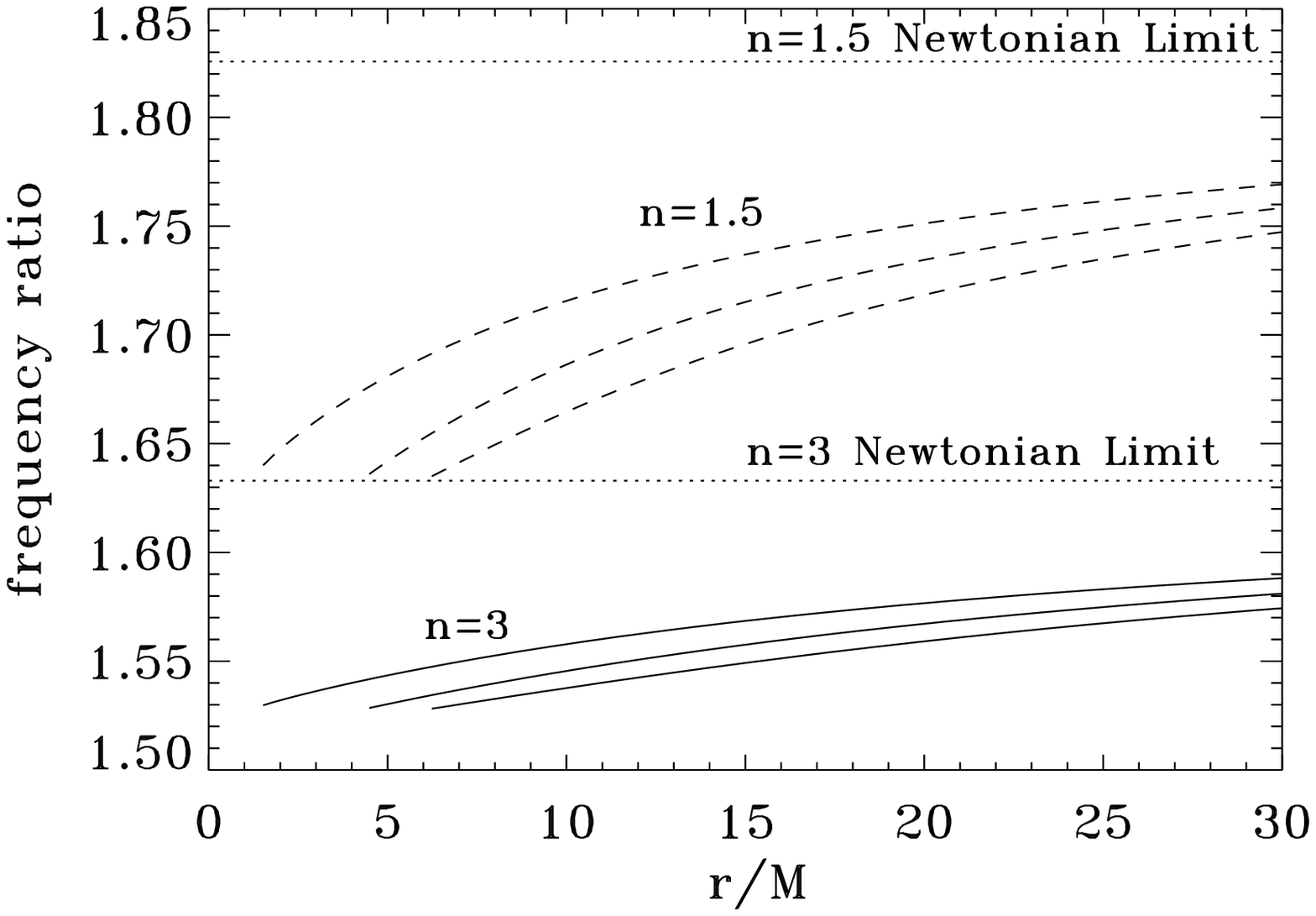}
 \caption{Ratio of the ``breathing mode'' acoustic wave $(+,+,j=1,k=0)$
frequency to the vertical epicyclic mode frequency, for a slender constant
specific angular momentum $(\kappa_0^2=0$) torus in a Kerr spacetime at
different Boyer-Lindquist coordinate radii.  Solid curves are for a  
polytropic index $n=3$, while dashed curves are for $n=1.5$.  From top to
bottom, each of the three curves at fixed $n$ corresponds to  
spin parameters $a/M=0$, 0.5, and 0.998, respectively.  The Newtonian limits
of the frequency ratios from equation (\ref{nujk}), $(8/3)^{1/2}$ for $n=3$
and $(10/3)^{1/2}$ for $n=1.5$, are shown by the horizontal dotted lines.
\label{figbreathmode}}
\end{figure}

\begin{figure}
 \includegraphics[width=160mm]{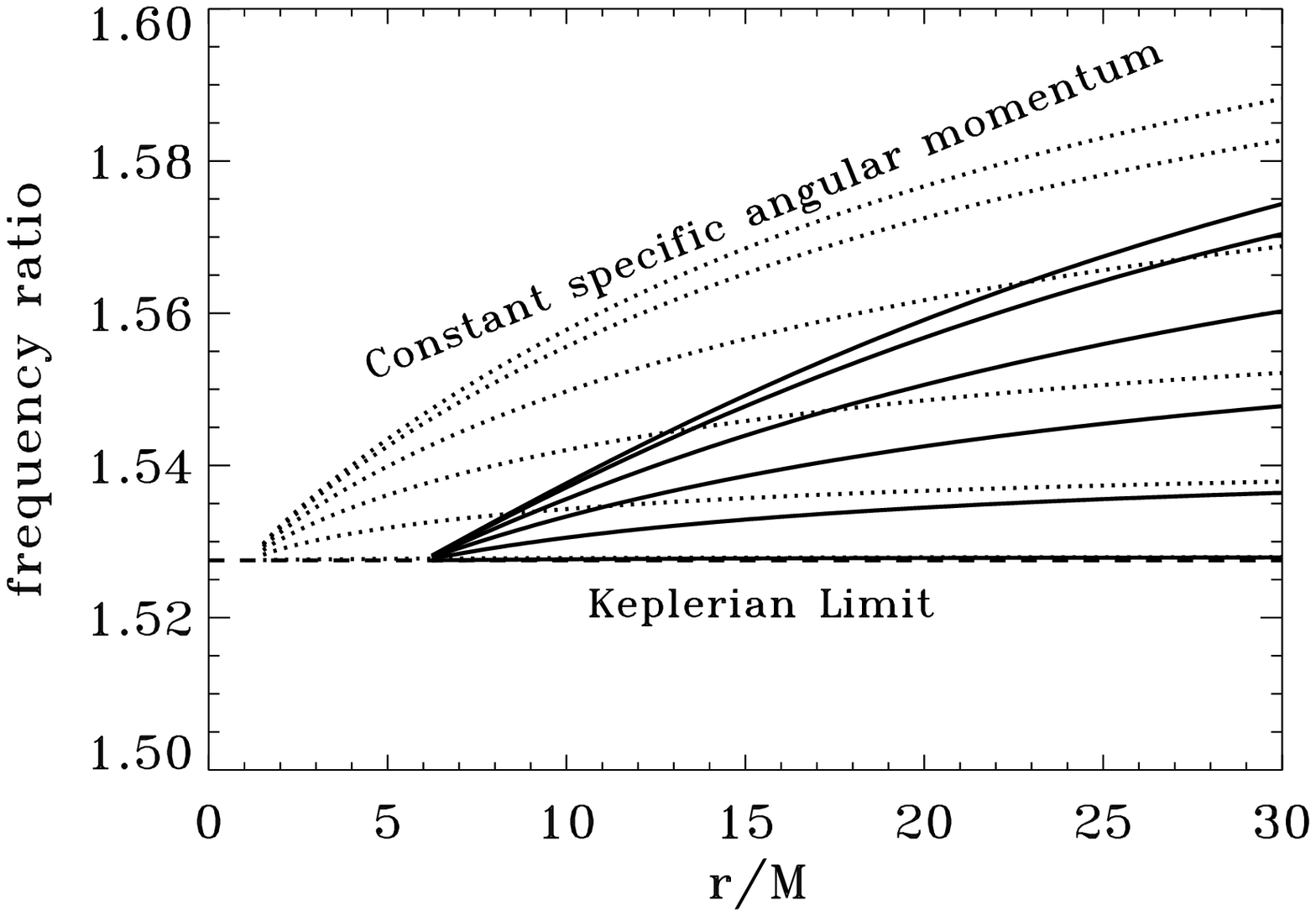}
 \caption{Ratio of the ``breathing mode'' acoustic wave $(+,+,j=0,k=2)$
frequency to the vertical epicyclic mode frequency, for a slender $n=3$ torus
in an $a/M=0.998$ Kerr spacetime (dotted curves) and a Schwarzschild spacetime
(solid curves) at different Boyer-Lindquist coordinate
radii and different values of $\kappa_0$.  From top to bottom, the curves
have $\kappa_0/\omega_r=0$, 0.2, 0.4, 0.6, 0.8, and 0.99.  The
$\kappa_0/\omega_r=0.99$ curves are of course close to the Keplerian limit
(eq. \ref{vertfreq} with $n_y=2$), shown as the horizontal dashed line.
\label{figbreathmoderkr}}
\end{figure}

Our general mode analysis reveals that there is another pair of modes whose
frequency ratio is not too far from 1.5:  the ``breathing mode'' acoustic
wave and the vertical epicyclic mode.  Figures \ref{figbreathmode} and
\ref{figbreathmoderkr} show the ratios of these two mode frequencies for a
variety of equilibrium parameters.  For $n=3$ tori in particular, which are
probably most relevant to radiation pressure supported configurations, the
frequency ratio is close to 1.5 even in the Keplerian limit.  This is in
contrast to the plus-mode/radial epicyclic mode pair, whose ratio goes to
unity in the Keplerian limit.  (The reader should compare Figures
\ref{figplusmoderkr} and \ref{figbreathmoderkr}, and note the different
scalings on the vertical axis.)

These are but just two examples of possible mode pairs which may be related
to the observed $3:2$ commensurability of high frequency QPO's in black
hole X-ray binaries.  There are presumably others.  In particular, we have
only considered axisymmetric ($m=0$) frequency ratios in this section,
but of course $m\ne0$ would allow an even richer set of possibilities.

\section{Comparison with Numerical Simulation}

A number of authors have performed axisymmetric numerical simulations of
hydrodynamic tori in order to explore their oscillation modes.  Different
techniques have been used to excite the oscillations.  For example,
\citet{zan05} tried initial radial velocity perturbations, initial density
perturbations, and perturbations based on the eigenfunctions of the modes
of vertically integrated tori.  \citet{lee04} and \citet{rub05a,rub05b}
have also tried periodic external forcing of the inner regions of the torus
at a specific frequency.

Here we use the analytic solutions from section 5 to
excite a spectrum of modes in slender hydrodynamic tori and then
allow them to evolve numerically for a large number of orbital
periods. The goals are to verify that we can indeed identify the
modes at subsequent stages of simulation and lay the foundation for
further numerical study of more general torus configurations.
The numerical evolution was carried out
using the {\it Cosmos++} relativistic magnetohydrodynamic code described
in \citet{ann05}.

The present simulations are restricted to two-dimensional
axisymmetry in order to expedite this work and avoid the growth of
non-axisymmetric instabilities \citep{pap84}. We use a single-level
fixed mesh with a resolution of $128^2$.  We use a Schwarzschild spacetime,
and the Schwarzschild coordinate grid covers an angular scale
$0.488\pi \le \theta \le 0.512\pi$ and has radial boundaries at
$r_{min}=9.7 M$ and $r_{max}=10.3 M$. The pressure maximum of the
unperturbed torus is located at $r_0=10.0 M$, while its inner
radius is $r_{in}=9.9 M$. The orbital period at the pressure maximum is
$\tau_{orb}=199 M$. The polytropic index is $n=3$ and the polytropic
constant is chosen such that the sound speed at the pressure maximum
is $c_{s,0}=0.0011$.  The specific angular momentum distribution is
chosen to be a power law,
$$
\ell=\ell_0\left({\lambda\over\lambda_0}\right)^{2-q},
$$
where, for Schwarzschild spacetime,
$\lambda\equiv(-g^{tt}/g^{\phi\phi})^{1/2}$.

We initially excite a set of modes by using the eigenfunctions of Tables
3 or 5.  The phase of the oscillation modes are initially chosen such that
the perturbations only need to be applied to the velocity fields and
not density or pressure.  Following \citet{zan05}, we tracked the modes by
looking at the Fourier transform of the norm of the rest-mass
density $||\rho||^2 \equiv \sum_{i=1}^{N_r} \sum_{j=1}^{N_\theta}
\rho_{ij}^2$ and making a comparison with the predicted mode
frequencies. As we show, this method successfully identifies many of the
desired modes, but does a poor job on others.  It also
reveals power spectrum peaks that do not correspond to the initial set of
modes that were excited.

\begin{figure}
\includegraphics[width=160mm]{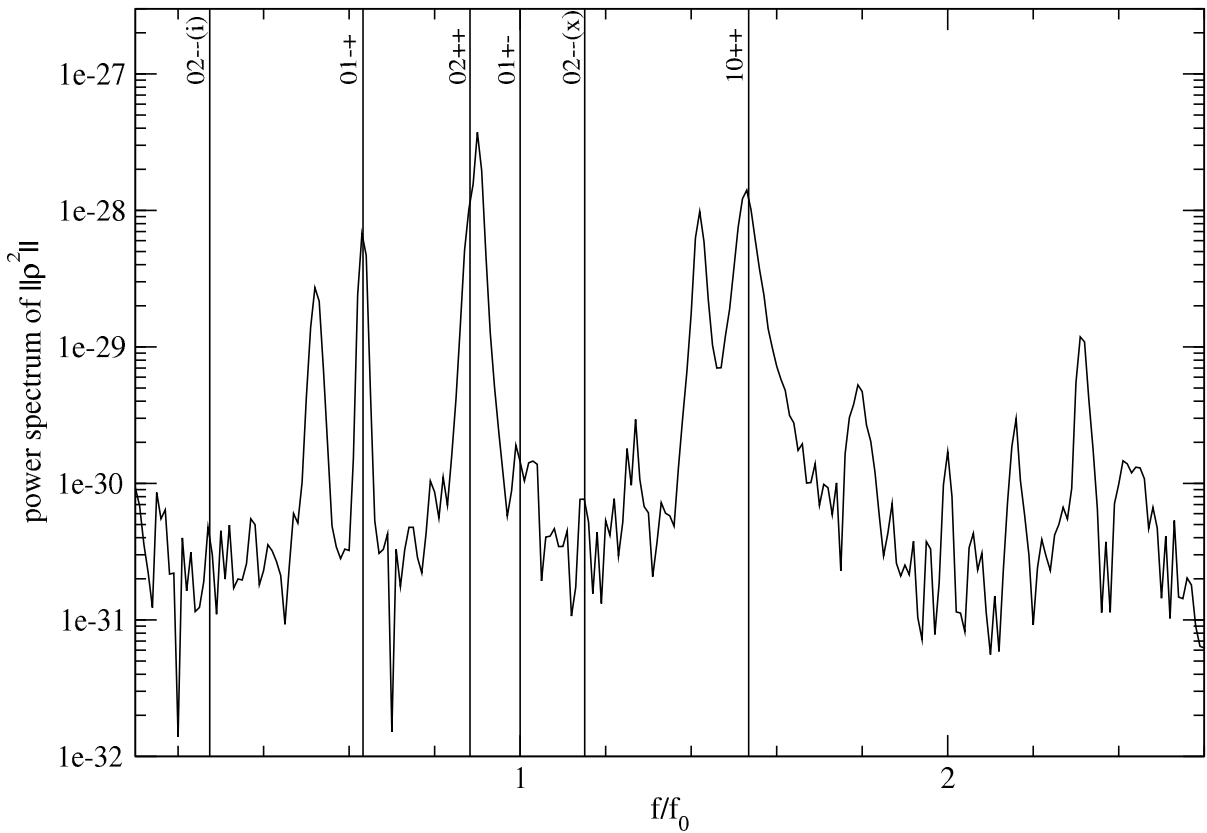}
\caption{Power spectrum of $||\rho||^2$ for a
torus with $\bar{\kappa}_0/\bar{\omega}_r=0.5$.  Each of the eigenfrequencies
from Table 2 are identified.  The units of the vertical axis are arbitrary,
while the horizontal axis has been scaled with the reciprocal of the orbital
period at the pressure maximum.
\label{fig:Fragile1}}
\end{figure}

\begin{figure}
\includegraphics[width=160mm]{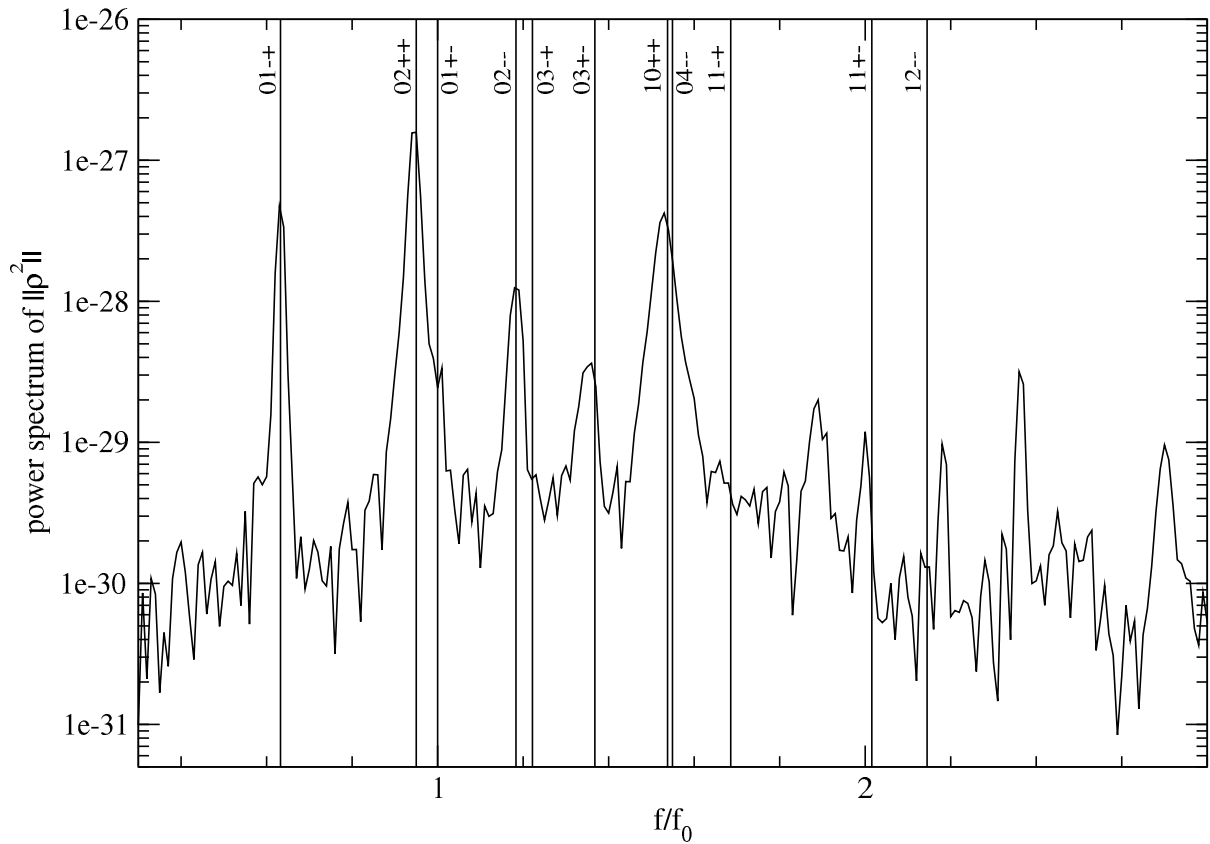}
\caption{Similar to Figure \ref{fig:Fragile1}, but for a constant specific
angular momentum torus with each of the eigenfrequencies from Table 4
identified.
\label{fig:Fragile2}}
\end{figure}

Figure \ref{fig:Fragile1} shows a portion of the power spectrum of
$||\rho||^2$ for a simulation of a non-constant specific angular
momentum torus ($l_0=3.95$ and $q=1.929$, so that
$\bar{\kappa}_0/\bar{\omega}_r=0.5$). Each of the eigenfrequencies
from Table 2 is labeled. Figure \ref{fig:Fragile2} shows
a similar power spectrum for a constant specific angular momentum
torus ($l=3.95$) with each of the eigenfrequencies from Table 4
indicated.  While many of the eigenfrequencies are clearly present in
the power spectrum, several are not.  In particular, the vertical
epicyclic mode (01+-) is not obvious, even though it was definitely
excited by the initial conditions.  We suspect that this is due to the fact
that this mode produces very little in terms of density changes, and so
the $||\rho||^2$ diagnostic is not sensitive to its presence.
(The fact that the radial epicyclic mode is still easily detected is
probably due to the squeezing of the torus by the increasing tidal field of
the black hole during the inward phase of the oscillation.)  We also tried
using $||\rho-\rho_i||^2$ as a diagnostic, where $\rho_i$ is the initial
density field of the torus, and found that the vertical epicyclic mode was
much stronger in the resulting power spectrum, while other modes were missing.
More work needs to be done to investigate more sensitive numerical diagnostics
for these modes.

The prominent peaks at $f/f_0\simeq 0.53, 1.40$ and $1.80$ in Figure
\ref{fig:Fragile1} are likely not true oscillation modes. The nonlinear
interaction of two waves with frequencies $f_1$ and $f_2$, if not forbidden
by selection rules (parity, etc.), will produce modulation at ``combination
frequencies" $|f_1\pm f_2|$. The three peaks may then be explained as
$f_{10++}-f_{01+-}=1.53-1.00\simeq 0.53$,
$0.53+f_{02++} = 0.53+0.88\simeq 1.41$, and
$2f_{02++} = 2\times 0.88 \simeq 1.76$, respectively.
Alternatively, we find numerically that there is a higher order even-even
parity mode with frequency $1.78$, implying a near resonance with $2f_{02++}$.
Lastly, we note that there is a chance near resonance
$f_{01-+} + f_{02++} \simeq f_{10++}$. In the slender torus, this interaction
is forbidden by the parity selection rule for the radial direction. However,
this parity rule is strictly valid only for the infinitely slender torus, and
may still occur in our numerical simulations.

\section{Discussion and Conclusions}

We have presented a rather complete discussion of the oscillation modes of
non-self-gravitating, relativistic, polytropic tori in an arbitrary axisymmetric
spacetime with reflection symmetry, in the limit where the tori are very
slender.  Non-slender torus oscillation modes can be explored numerically,
and some work on this has already been done by \citet{zan05} and
\citet{rub05b}.  It would also
be possible to examine the behavior of modes for thicker tori analytically
using perturbative methods, generalizing the technique employed by
\citet{bla85}.

All the slender tori considered in this paper are hydrodynamically unstable
to the nonaxisymmetric, Papaloizou-Pringle instability (PPI) \citep{pap84}.
They would also develop magnetohydrodynamical turbulence in the presence of a
weak magnetic field because of the MRI \citep{bal98}.  Almost all simulations
examining oscillations of tori, including those presented in this paper,
have been axisymmetric and neglected magnetic fields, so both of these
instabilities have not been allowed to develop.  A major outstanding issue
is therefore which, if any, of the hydrodynamic torus modes survive these
instabilities.  A preliminary investigation of this has been done by
\citet{fra05}, but more work needs to be done.  At least in radially extended
tori, the PPI can be suppressed
by dynamical accretion through the torus inner edge \citep{bla87,dev02}.
Moreover, the MRI appears to dominate the PPI when magnetic fields are
included, and generally leads to a near-Keplerian distribution of
specific angular momentum in the flow \citep{haw02}.  If modes which are
essentially hydrodynamic in nature can exist in the presence of MRI turbulence,
then the near Keplerian limit is therefore probably of most interest.  The
turbulence itself may then determine the mode amplitudes, but all of these
questions remain to be investigated.  A systematic understanding of all the
types of modes which may be present, as we have done in this paper, should
aid such investigations.

Figure \ref{fig:sigmavskappa} shows that all mode families converge to the
epicyclic frequencies
or vertical sound waves in the Keplerian limit. If physical discs are
indeed nearly Keplerian, this greatly simplifies seismology as only
a few distinct frequencies exist. But how close are real discs to the
Keplerian limit? Equation (8) can be rewritten in terms of the angular
momentum gradient for Keplerian motion, $\partial l_K/\partial r$, by
differentiating equation (6) and substituting into equation (8). The result
is exactly the same as for $\kappa$ in equation (16), with $\partial
l/\partial r$, the angular momentum gradient for the fluid torus,
replaced by $\partial l_K/\partial r$.  The difference of these two
quantities becomes
\begin{equation}
\omega_r^2-\kappa_0^2 = \frac{E_0^2}{l_0A_0^2} 
\left( \frac{g^{tt}_{,r}-l g^{t\phi}_{,r}}{g_{rr}} \right)_0
\frac{\partial}{\partial r}\left( l_K - l \right),
\label{eq:diff}
\end{equation}
which manifestly goes to zero in the Keplerian limit.

\citet{dev03} present plots of $l_K$ and $l$ (averaged
against density over spherical shells) against radius for MRI driven
accretion onto Kerr black holes. In the region of the ``inner torus"
(see their Figure 3), they find $l$ has a similar slope as $l_K$, but is
smaller by $\sim 1-10\%$, the difference going to zero at the marginally
stable orbit. Using equation \ref{eq:diff}, this implies $(\omega_r^2-
\kappa_0^2)/\omega_r^2 \sim 0.01-0.1$ in our notation. Hence Figure 2
shows that low order modes should be quite accurately described by the
Keplerian limit, while the deviation from the Keplerian limit increases
with number of nodes. For emission from an optically thick region, flux
variations from the photosphere become smaller as more angular nodes
are included due to averaging of hot and cold spots. Hence low order
modes, which are more nearly in the Keplerian limit, are also likely
more observable.

\begin{figure}
\includegraphics[width=160mm]{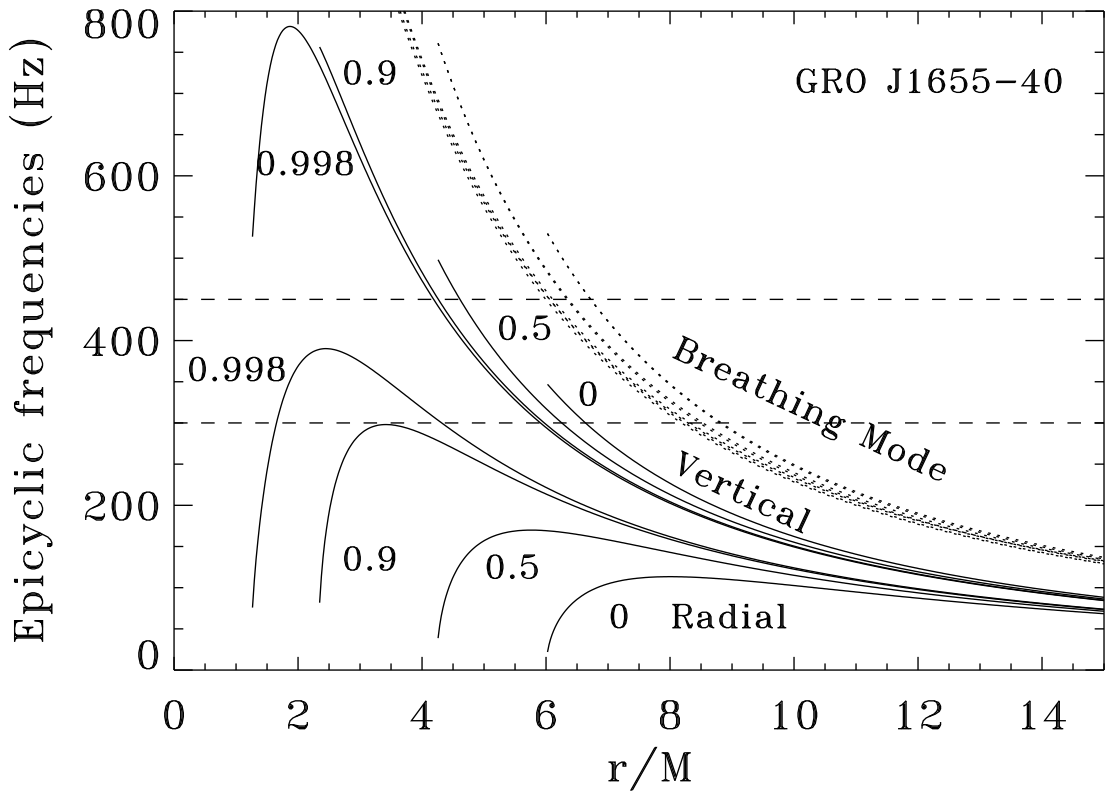}
\caption{Vertical (upper set of curves) and radial (lower set of curves)
epicyclic mode frequencies as a function of Boyer-Lindquist coordinate radius
for a black hole of mass 6.3~M$_\odot$, appropriate for GRO J1655-40, and
various black hole spins $a/M$ as labelled.  The horizontal dashed lines
indicate the frequencies of the observed high frequency QPO's in this source.
Models which identify the 300 Hz QPO as the axisymmetric radial epicyclic
mode require $a/M\gta0.9$.  Dotted lines show the lowest order breathing mode
frequency for the same black hole spins and both constant specific angular
momentum tori and tori in the Keplerian limit.  Models which identify the
300 Hz QPO with the vertical epicyclic mode and the 450 Hz QPO with the
breathing mode can work for any spin.
\label{fig:nuepi1655}}
\end{figure}

In view of this, the lowest order breathing mode/vertical acoustic wave and
the vertical epicyclic mode may be an attractive candidate for a pair of
modes which have a near 3:2 commensurability for radiation pressure-dominated
($n=3$) tori, as this persists in the Keplerian
limit.  This is in contrast to the plus mode and radial epicyclic mode pair,
whose frequency ratio approaches unity in the Keplerian limit.  Identifying
the vertical epicyclic mode with the lower of the two observed frequencies
also allows the black holes to have any spin parameter, in contrast to the
case where the lower of the two frequencies is identified as the radial
epicyclic mode.  This is illustrated in Figure \ref{fig:nuepi1655} for
the case of GRO J1655-40.  This source has a mass of $\sim 6.3$~M$_\odot$
\citep{oro03}, and models which identify the 300~Hz QPO as the axisymmetric
radial epicyclic mode require $a/M\gta0.9$ \citep{rez03a,tor05}.  The
same is true for diskoseismology models which identify the 300~Hz QPO with an
axisymmetric ``$g$-mode'', which has a frequency below the radial epicyclic
frequency within the mode trapping region (e.g. \citealt{wag01}).  This is
in conflict with recent continuum spectral fitting results which suggest
a lower black hole spin \citep{sha05}.  If the 300~Hz QPO is instead the
axisymmetric vertical epicyclic mode, any spin is possible.  Many other
mode pairs would probably also work, especially if we consider nonaxisymmetric
($m\ne0$) modes.  However, one still needs to explain why the QPO's
occur at the actual observed frequencies, i.e. what singles out a preferred
radius for the torus in this source.

More work needs to be done to
investigate how torus oscillation modes might manifest themselves in X-ray
light curves and power spectra.  Some work has been done based on tracking
photon geodesics from the torus to the observer \citep{bur04,sch05}, but
additional work needs to be done on the photon emission mechanism of the
tori themselves.  The difficulties that some diagnostics have in identifying
torus oscillation modes in numerical simulations, even when they are present
in the initial conditions as discussed in section 6 above, should serve
as a warning to numericists searching for QPO's in their simulations.
The X-ray observations generally have nothing to do with these diagnostics.

It remains unclear whether tori are appropriate models for the flow geometry
for the ``steep power law state'' \citep{mcc05} where high frequency QPO's
are seen in black hole X-ray binaries.  \citet{kub04} suggest that the
steep power law state consists of a standard geometrically thin disc sandwiched
between a magnetized corona which has sucked up much of the accretion power
out of the disc.  Oscillation modes that are analogous to the torus modes
with vertical reflection symmetry {\it might} exist in such a magnetized
corona, because the vertical velocities vanish in the equatorial plane, where
the high inertia disc is located.  We hope that our comprehensive hydrodynamic
mode analysis may act as a foundation for investigation of these and the other
questions noted above.

\section*{Acknowledgments}

We thank Marek Abramowicz, W{\l}odek Klu\'zniak, Luciano Rezzolla,
Shin Yoshida, and Olindo Zanotti for very useful discussions.  This
research was supported in
part by the National Science Foundation under grants PHY99-0794 and
AST03-07657, and under the following NSF programs: Partnerships for
Advanced Computational Infrastructure, Distributed Terascale Facility (DTF)
and Terascale Extensions: Enhancements to the Extensible Terascale Facility.

\label{lastpage}

\end{document}